\documentclass[a4paper,onecolumn,accepted=2026-02-24]{quantumarticle}
\pdfoutput=1
\usepackage{orcidlink}

\usepackage{amsthm}
\usepackage{amsmath,bm}
\usepackage{amssymb}
\usepackage{amsfonts}
\usepackage{graphicx}
\usepackage{txfonts}
\usepackage{xcolor}
\usepackage{float}
\usepackage{braket}
\usepackage[capitalise]{cleveref}
\usepackage{bbm}
\usepackage{changes}

\PassOptionsToPackage{sort&compress}{natbib}
\usepackage[numbers]{natbib}

\newcommand{\abs}[1]{\ensuremath{\left|{#1}\right|}}

\newcommand{\var}{\text{Var}} 
\newcommand{\dd}{\text{d}}

\newcommand{\ie}{\textit{i.e.} }
\newcommand{\eg}{\textit{e.g. }}

\newcounter{SaveEqnCntr}
\newcommand\SaveEqCtr{\setcounter{SaveEqnCntr}{\value{equation}}}
\newcommand\RestoreEqCtr{\setcounter{equation}{\value{SaveEqnCntr}}}

\begin{document}

\title{Quantum entanglement in phase space}

\author{Shuheng Liu}
\thanks{These authors contributed equally to this work.}
\address{State Key Laboratory for Artificial Microstructure and Mesoscopic Physics, School of Physics, Frontiers Science Center for Nano-optoelectronics, $\&$ Collaborative Innovation Center of Quantum Matter, Peking University, Beijing 100871, China}

\author{Jiajie Guo}
\thanks{These authors contributed equally to this work.}
\address{State Key Laboratory for Artificial Microstructure and Mesoscopic Physics, School of Physics, Frontiers Science Center for Nano-optoelectronics, $\&$ Collaborative Innovation Center of Quantum Matter, Peking University, Beijing 100871, China}

\author{Qiongyi He}
\email{qiongyihe@pku.edu.cn}
\address{State Key Laboratory for Artificial Microstructure and Mesoscopic Physics, School of Physics, Frontiers Science Center for Nano-optoelectronics, $\&$ Collaborative Innovation Center of Quantum Matter, Peking University, Beijing 100871, China}
\address{Collaborative Innovation Center of Extreme Optics, Shanxi University, Taiyuan, Shanxi 030006, China}

\author{Matteo Fadel}
\email{fadelm@phys.ethz.ch}
\address{Department of Physics, ETH Z\"{urich}, 8093 Z\"{urich}, Switzerland}

\begin{abstract}
While commonly used entanglement criteria for continuous variable systems are based on quadrature measurements, here we study entanglement detection from measurements of the Wigner function. 
These are routinely performed in platforms such as trapped ions and circuit QED, where homodyne measurements are difficult to be implemented. 
We provide complementary criteria which we show to be tight for a variety of experimentally relevant Gaussian and non-Gaussian states.
Our results show novel approaches to detect entanglement in continuous variable systems and shed light on interesting connections between known criteria and the Wigner function.
\end{abstract}

\maketitle

\section*{Introduction}
Most criteria designed to detect entanglement in experiments with continuous variable (CV) systems rely on quadrature measurements. A primary example is the Simon \cite{SimonPRL2000} and Duan \textit{et al.} \cite{DuanPRL} criteria, which state that for all separable states 
\begin{equation}\label{duan}
\var[\hat{x}_A+\hat{x}_B] + \var[\hat{p}_A-\hat{p}_B] \geq 4  \;,
\end{equation} 
where $\hat{x}_k$ and $\hat{p}_k$ are respectively the position and momentum quadratures of system $k \in \{A,B\}$ satisfying $[\hat{x}_k,\hat{p}_l]=2i \delta_{kl}$. Observing a violation of Eq.~\eqref{duan} thus implies entanglement between $A$ and $B$. Moreover, a simple generalization of this condition has been shown to be necessary and sufficient for detecting entanglement in Gaussian states \cite{DuanPRL}.

\begin{figure}[t]
    \begin{center}
	\includegraphics[width=0.55\columnwidth]{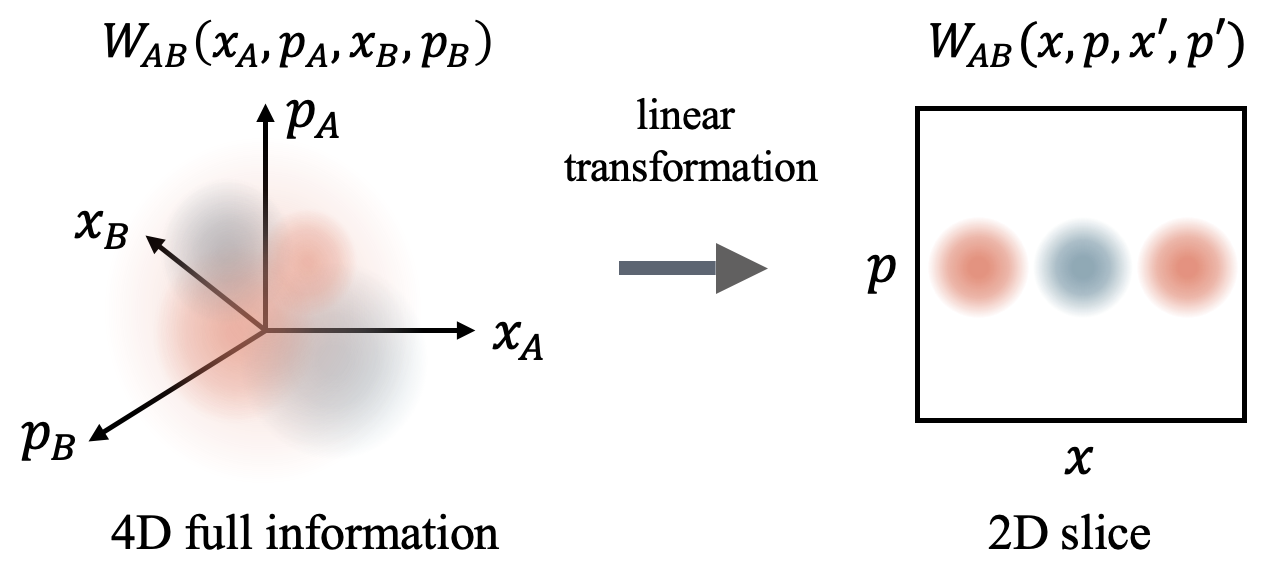}
	\end{center}
    \caption{Full information of a bipartite continuous-variable system is contained in a joint Wigner function $W_{AB}(x_A,p_A,x_B,p_B)$ defined in a four-dimensional phase space. Revealing entanglement from measurements of this function is in general a difficult task. To overcome this problem, we present entanglement criteria based on measurements on a two-dimensional slice $W_{AB}(x,p,x',p')$, where $x'$ and $p'$ are linear functions of $x,p$, representing a coordinate transformation.}
    \label{Fig1_Illustration}
\end{figure}

Besides propagating optical and microwave fields where homodyne measurements are routinely implemented, in a wide range of other CV systems is lacking the possibility of performing quadrature measurements in a natural way. Examples include the motional states of trapped ions~\cite{QuantumLeibfriedRMP2003}, the microwave cavity modes in cavity~\cite{ManipulatingRaimondRMP2001} and circuit quantum electrodynamics~\cite{CircuitBlaisRMP2021} and the vibration modes in circuit quantum acoustodynamics \cite{UweNP2022}.
In these platforms, which are described by the Jaynes-Cummings model, a much more natural approach is to perform measurements of the Wigner function in phase space through displaced parity measurements \cite{LutterbachPRL97}.

It is thus natural to ask whether entanglement between two systems can be concluded directly from measurements of their joint Wigner function. Formally, it is known that for all separable states $\rho_{AB} = \sum_\lambda p(\lambda) \rho_A^{\lambda} \otimes \rho_B^{\lambda}$ the joint Wigner function factorizes as
\begin{equation}\label{eq:SeparableWigner}
W_{A B}\left(x_A, p_A, x_B, p_B\right)=\sum_\lambda p(\lambda) W_A^\lambda\left(x_A, p_A\right) W_B^\lambda\left(x_B, p_B\right) \;,
\end{equation}
but testing this condition is extremely impractical. To be of convenient experimental implementation, one would like an entanglement witness involving only a few measurements of the Wigner function.

Known examples in this direction are the Bell inequality by Banaszek and Wódkiewicz~\cite{BanaszekPRL1999,BanaszekPRA1998}, which allows for the detection of nonlocality from parity measurements at only four points in phase-space, and the Einstein-Podolsky-Rosen (EPR) criterion by Walschaers and Treps~\cite{WalschaersPRL20}, which allows for the detection of steering when a state with negative Wigner function is remotely prepared. These criteria, however, are intrinsically much more demanding, since revealing nonlocality and steering requires much lower levels of noise compared to entanglement. Recently, some works pointed out that performing a balanced beam-splitter transformation on $\rho_{AB}$ allows for revealing entanglement properties from the reduced Wigner function of one of the transformed modes \cite{DynamicsJayachandran2023,CertifiableZaw2024}. These results suggest the possibility of deriving entanglement from a few measurements in phase-space, without the need of full tomography.

In this work, we study entanglement detection from the Wigner function and show that for all separable states
\SaveEqCtr 
\begingroup 
\setcounter{equation}{0}
\renewcommand\theequation{\Roman{equation}} 
\renewcommand\theHequation{AA\arabic{equation}} 
\begin{gather}
\iint_{-\infty}^{\infty}  W_{AB}(x\cos\theta, p\cos\theta, x'\sin\theta, p'\sin\theta) \,\dd x \dd p \leq \frac{1}{2\pi} \label{crit1}\\
\iint_{R} \left|W_{A B}\left(x \cos \theta, p \cos \theta, x^{\prime} \sin \theta, p^{\prime} \sin \theta\right)\right| \dd x \dd p \leq \frac{1}{2 \pi |\sin 2\theta|}  \label{crit2} \\
\iint_{-\infty}^{\infty}  W_{AB}(x, p, x', p') \,\dd x \dd p \geq 0  \label{crit3}
\end{gather}
\endgroup
\RestoreEqCtr 
where $[x'=x'(x,p),p'=p'(x,p)]$ is an arbitrary linear transformation of the coordinate system for $B$, see Eq.~\eqref{eq:QuadratureTransformation} and Fig.~\ref{Fig1_Illustration}, and $R$ is an arbitrary region in phase space. The condition $\sin 2\theta \neq 0$ is assumed.
We provide a method to derive the optimal transformation and show that known criteria \cite{SimonPRL2000,DuanPRL,DynamicsJayachandran2023,CertifiableZaw2024,ZhangDetecting2013} can be understood as specific choices.
In particular, we prove that for Gaussian states criterion~\eqref{crit1} is necessary and sufficient, as it coincides with Duan-Simon criteria, and that criterion~\eqref{crit3} can be expressed as the measurement of the reduced Wigner function at a single point in the phase space of a transformed basis.
We then show examples demonstrating how our criteria successfully detect entanglement in a variety of routinely prepared Gaussian and non-Gaussian CV states, such as two-mode squeezed states, Werner states and entangled cat states.

\section*{Results}

\vspace{2mm}
\textbf{Wigner function formalism.}-- Consider a system of $N$ bosonic modes, each described by creation and annihilation operators $\hat{a}_k^{\dagger}$ and $\hat{a}_k$, which satisfy the commutation relation $\left[\hat{a}_k, \hat{a}_{k^{\prime}}^{\dagger}\right]=\delta_{k k^{\prime}}$ for $k,k^\prime\in\{1,\dots,N\}$. For each mode, it is natural to introduce position and momentum operators $\hat{x}_k=\hat{a}_k^{\dagger}+\hat{a}_k$ and $\hat{p}_k=i\left(\hat{a}_k^{\dagger}-\hat{a}_k\right)$, satisfying $[\hat{x}_k,\hat{p}_{k^\prime}]=2i\delta_{k k^{\prime}}$. 
We arrange these in the vector $\hat{\boldsymbol{\xi}}=\left(\hat{x}_1, \hat{p}_1, \ldots, \hat{x}_N, \hat{p}_N\right)^{\top}$, which is associated to the vector of phase space coordinates $\boldsymbol{\xi} \equiv\left(x_1, p_1, \ldots, x_N, p_N\right)^{\top}$. A convenient representation of the state $\rho$ of a CV system is given by the Wigner function
\begin{equation}
W(\boldsymbol{\xi})=\int_{\mathbb{R}^{2 N}} \frac{d^{2 N} \boldsymbol{\alpha}}{(2 \pi)^{2 N}} \exp \left(-i \boldsymbol{\xi}^{\top} \boldsymbol{\Omega} \boldsymbol{\alpha}\right) \chi(\boldsymbol{\alpha}) \;,
\end{equation}
where $\boldsymbol{\Omega}=\bigoplus_1^N  \bigl( \begin{smallmatrix}0 & 1\\ -1 & 0\end{smallmatrix}\bigr)$ is the symplectic form and $\chi(\boldsymbol{\alpha})=\operatorname{Tr}\left[\hat{\rho} \exp \left(i \hat{\boldsymbol{\xi}}^{\top} \boldsymbol{\Omega} \boldsymbol{\alpha}\right)\right]$ is the characteristic function. 
Crucially, the value of the Wigner function at one point in phase space can be directly measured experimentally by using the fact that it can be written as a displaced parity measurement~\cite{DisplacedBishop1994,SeriesMoyaCessa1993,WignerTilma2016}, namely
\begin{equation}
W(\boldsymbol{\xi}) = \left(\dfrac{1}{2\pi}\right)^N \operatorname{Tr}\left[D(\boldsymbol{\xi}) \hat{\Pi} D^\dagger(\boldsymbol{\xi}) \hat{\rho} \right]\;,
\end{equation}
where $D(\boldsymbol{\xi})$ is the displacement operator and $\hat{\Pi}=(-1)^{\hat{n}_1+...+\hat{n}_N}$ is the joint parity operator. Direct Wigner function measurements are routinely performed on $N=1$ mode in a variety of platforms \cite{QuantumLeibfriedRMP2003,ManipulatingRaimondRMP2001,CircuitBlaisRMP2021,UweNP2022}, and they have also been demonstrated for $N=2$ modes in circuit QED \cite{ChenScience2016}.

The rest of this work will deal with a bipartite, \ie $N=2$, system described by the density matrix $\rho_{AB}$ or, equivalently, by the joint Wigner function $W_{AB}(x_A,p_A,x_B,p_B)$. Our goal is to detect entanglement between the $A$ and $B$ subsystems based on this function. Crucially, as illustrated in Fig.~\ref{Fig1_Illustration}, our criteria do not require characterizing the entire four-dimensional Wigner function. Instead, they rely on specific two-dimensional slices where the variables $(x_B, p_B)$ are constrained to be functions of $(x_A, p_A)$, reducing the measurement complexity from four independent variables to just two. (For notational simplicity, we then drop the subscript $A$.) Note that such a result is also useful towards revealing multipartite entanglement, as one can test inseparability across all possible bipartitions of the $N>2$ subsystems.

The Wigner function possesses several important properties that will be useful to derive our main results. For single mode states $\rho_A,\rho_B$, the trace of their product can be expressed as $\operatorname{Tr}\left[\rho_A \rho_B\right]=4 \pi \iint_{-\infty}^{\infty} W_A(x,p) W_B(x,p) \dd x \dd p$ which is nonnegative and often used to represent the fidelity to a pure state~\cite{FerraroGaussian2005}. This leads to
\begin{equation}\label{eq:WignerTraceProductBound}
\iint_{-\infty}^{\infty} W_A(x,p) W_B(x,p) \dd x \dd p \geqslant 0.
\end{equation}
When $\rho_A=\rho_B=\rho$, it represents the purity of the state $\rho$, which leads to
\begin{equation}\label{eq:WignerPurityBound}
\frac{1}{4 \pi} \geq \iint_{-\infty}^{\infty} [W(x, p)]^2 \dd x \dd p > 0 \;.
\end{equation}

A valid Wigner function remains such even under certain transformations of the phase space coordinate system~\cite{SimonPRL2000,becker2021convergence,RevzenBell2005}.
Specifically, we consider the following linear transformation on the phase space coordinates $x$ and $p$,
\begin{equation}\label{eq:QuadratureTransformation}
\left(\begin{array}{l}
x^{\prime}\\p^{\prime}
\end{array}\right)
=\left(\begin{array}{ll}
a & b \\
c & d
\end{array}\right)
\left(\begin{array}{l}
x\\p
\end{array}\right)
+\left(\begin{array}{l}
x_0\\p_0
\end{array}\right)
\end{equation}
with determinant $\Delta = a d-b c=\pm 1$. This transformation can represent the combination of a canonical coordinate transformation and a mirror reflection in phase space. 

A determinant $\Delta=+1$ indicates a canonical transformation preserving the commutator $[x^{\prime},p^{\prime}]=[x,p]=i\hbar$. Such a transformation of the quadratures $(x, p)$ is often referred to as a symplectic transformation, where $(x, p)$ and $(x^{\prime}, p^{\prime})$ are related through a unitary transformation, $x^{\prime}=U^{\dagger} x U, p^{\prime}=U^{\dagger} p U$, as their spectra are preserved up to a constant displacement $(x_0,p_0)$~\cite{RevzenBell2005}. To transition from the Heisenberg picture to the Schrödinger picture, we move from a unitary transformation of the operator to a unitary transformation of the quantum state. This unitary transformation affects the Wigner function as $W_{U \rho U^{\dagger}}(x^{\prime}, p^{\prime})=W_\rho\left(x,p\right)$. Typical unitary transformations are displacement, rotation and single mode squeezing.

On the other hand, a determinant $\Delta=-1$ indicates a mirror reflection given by $\left(\begin{smallmatrix}
\cos(2 \vartheta) & \sin(2 \vartheta) \\
\sin(2 \vartheta) & -\cos(2 \vartheta)
\end{smallmatrix}\right)$. This reflection occurs along the axis which is rotated counterclockwise by an angle $\vartheta$ relative to the positive $x$-axis. Such transformation is associated to an anti-unitary operator, which leaves the absolute value of the inner product $|\langle x, y\rangle|$ invariant according to Wigner's theorem~\cite{wigner2012group}. As an example, time reversal $(x,p)\rightarrow (x,-p)$ is a typical anti-unitary operator.

To summarize, the local linear coordinate transformation Eq.~\eqref{eq:QuadratureTransformation} corresponds to a change of the state $f(\rho)\rightarrow \rho$ resulting in
\begin{equation} \label{eq:Mtransform}
W_{f(\rho)}(x, p)=W_\rho\left(x', p'\right) \;,
\end{equation}
which is still a valid Wigner function. In the SM Sec.~II~\cite{supplement}, we discuss the choices of the determinant $\Delta=\pm1$ for different criteria. Criterion~\eqref{crit1} is nontrivial when $\Delta=-1$ and Criterion~\eqref{crit3} is nontrivial when $\Delta=1$.

\vspace{2mm}
\textbf{Entanglement in phase space.}-- From the result Eq.~\eqref{eq:Mtransform} we note that applying the specified local transformation to the Wigner function of a separable state Eq.(\ref{eq:SeparableWigner}), namely doing
\begin{equation}
W_{A B}\left(x_A, p_A, x'_B, p'_B\right) = \sum_\lambda p(\lambda) W_A^\lambda\left(x_A, p_A\right) W_B^\lambda\left(x'_B, p'_B\right) \;,
\end{equation}
still results in a valid Wigner function. Without loss of generality, we let the local transformation apply only on mode $B$. 

For example, taking the axis of the mirror reflection as the $x$-axis, such a transformation has already been used to generalize the Peres-Horodecki separability criterion~\cite{PeresPRL1996,HorokechiPLA1997} to continuous variable systems~\cite{SimonPRL2000}, since the transpose operation of the density matrix $\hat{\rho} \rightarrow \hat{\rho}^T$ is equivalent to the mirror reflection $W(x,p) \rightarrow W(x,-p)$ in phase space. 
The Simon criterion detects entanglement by showing that the partially transposed state violates the uncertainty relation that separable states must satisfy~\cite{SimonPRL2000}. 
For two-mode Gaussian states, when the Simon criterion is violated and the partially transposed state is sent through a beam splitter, one output mode exhibits an extremely localized Wigner function that would correspond to an unphysical purity exceeding unity, as detailed in SM Sec.~III~\cite{supplement}. 
In contrast, our criterion \eqref{crit1} employs a fundamentally different approach: rather than testing uncertainty relation violations, it detects when the partially transposed state violates the upper bound of the Wigner function that all physical states must satisfy, in closer analogy to the original PPT criterion. 
This distinction, which we illustrate through a phase-space interpretation in SM Sec.~III~\cite{supplement}, explains why criterion \eqref{crit1} can outperform the Simon criterion for certain non-Gaussian states (see Fig.~\ref{Fig:SMtable} later in the text for a detailed comparison). 
Furthermore, for Gaussian states, the Simon criterion essentially utilizes the complete covariance matrix information, which fully determines the shape of the joint Wigner function (the first-order moments only affect the position of the Gaussian distribution). 
In contrast, criterion \eqref{crit1} relies solely on partial information from a specific phase-space slice, requiring less data than that needed for full reconstruction of the joint Wigner function, while still enabling reliable entanglement detection.

Apart from mirror reflection along the $x$-axis, the criteria~\eqref{crit1}-\eqref{crit3} we present are more general, since they consider any linear transformation $(x,p)\rightarrow (x',p')$ of the coordinate system that follows Eq.~\eqref{eq:QuadratureTransformation}. In addition, our criteria do not require full information of the state. We give a proof of these criteria in the following. We further show that criterion~\eqref{crit1} is necessary and sufficient for Gaussian states in the Sec.~V of SM~\cite{supplement}.

The key insight behind criterion~\eqref{crit1} is that if the initial state is separable, then applying partial transposition followed by a beam splitter operation should yield a physically valid state. The resulting Wigner function must satisfy the upper bound of any physical Wigner function. When this bound is violated, it reveals that our separability assumption was incorrect.

\vspace{2mm}
\textit{Proof of criterion~\eqref{crit1}} -- Consider an arbitrary transformation $(x',p')$ which can be determined by the real parameters $a, b, c, d, x_0, p_0$. Based on these parameters, we then construct a new single-mode transformation
$
h: (x, p)^T \rightarrow \left( \begin{smallmatrix}
  a & b\\
  c & d
\end{smallmatrix} \right) \left(
  x + \frac{X}{\cos \theta},
  p + \frac{P}{\cos \theta}
\right)^T + \left( 
  x_0 \sin \theta ,
  p_0 \sin \theta  \right)^T
$
where $X,P$ are arbitrary constant coordinates. 
Then according to Eq.~\eqref{eq:Mtransform} we obtain
\begin{equation}
\begin{aligned}
& W_{h(B)}^\lambda\left(x \sin \theta-\frac{X}{\cos \theta}, p \sin \theta-\frac{P}{\cos \theta}\right) \\
=& W_B^\lambda\left(\left(a x+b p+x_0\right) \sin \theta,\left(c x+d p+p_0\right) \sin \theta\right) \\
=& W_B^\lambda\left(x^{\prime} \sin \theta, p^{\prime} \sin \theta\right).
\end{aligned}
\end{equation}
Assume $\rho_{AB}$ is a separable state with a Wigner function given by Eq.~\eqref{eq:SeparableWigner}. The left-hand side of criterion~\eqref{crit1} can be rewritten as
\begin{equation}\label{eq:BeamSplitterTransformationWithTheta}
\begin{aligned}
& \iint_{-\infty}^{\infty} \sum_\lambda p(\lambda) W_A^\lambda(x\cos \theta , p\cos \theta ) W_{h\left(B\right)}^\lambda\left(x\sin \theta -\frac{X}{\cos \theta}, p\sin \theta -\frac{P}{\cos \theta}\right) \dd x \dd p \\
= & \iint_{-\infty}^{\infty} \sum_\lambda p(\lambda) W_A^\lambda\left(x\cos \theta +X\sin \theta , p\cos \theta +P\sin \theta \right) W_{h\left(B\right)}^\lambda\left(x\sin \theta -X\cos \theta , p\sin \theta -P\cos \theta \right) \dd x \dd p \;,
\end{aligned}
\end{equation}
where in the last row we substituted $x\rightarrow x+ X\tan\theta$ and similarly for $p$.
This can be interpreted as follows: Perform a beam-splitter transformation $\left(\begin{smallmatrix}
\cos \theta & \sin \theta \\ \sin \theta & -\cos \theta\end{smallmatrix}\right)$ to the two modes $A$ and $h(B)$, and denote the resulting modes as $A':(x,p)$ and $B':(X,P)$. Notice that Eq.~\eqref{eq:BeamSplitterTransformationWithTheta} is exactly the Wigner function of mode $B'$, with mode $A'$ integrated out. Therefore Eq.~\eqref{eq:BeamSplitterTransformationWithTheta} should be upper-bounded by $1/(2\pi)$~\cite{SchleichQuantumBook2001}. Detailed proof can be found in Sec.~IV of SM~\cite{supplement}.  $\Box$

\vspace{2mm}
Criterion~\eqref{crit2} exploits the fact that for separable states, the amount of correlation (with or without partial transposition) is fundamentally limited by the purities of the reduced states through the Cauchy-Schwarz inequality. When the measured correlations exceed this bound, they cannot be explained by classical correlations alone, thereby proving entanglement.

\vspace{2mm}
\textit{Proof of criterion~\eqref{crit2}} -- Using the expression given by Eq.~\eqref{eq:SeparableWigner}, we have for any separable state
\begin{align}
& \iint_{R}\left|W_{A B}\left(x \cos \theta, p \cos \theta, x^{\prime} \sin \theta, p^{\prime} \sin \theta\right)\right| \dd x \dd p  \notag\\
& =\iint_{R}\left|\sum_\lambda p(\lambda) W_A^\lambda(x \cos \theta, p \cos \theta) W_B^\lambda\left(x^{\prime} \sin \theta, p^{\prime} \sin \theta\right)\right| \dd x \dd p  \notag\\
& \leqslant \sum_\lambda p(\lambda) \iint_{R}\left|W_A^\lambda(x \cos \theta, p \cos \theta) W_{f_{\theta}(\rho_B^\lambda)}^\lambda(x \sin \theta, p \sin \theta)\right| \dd x \dd p  \notag\\
& \leqslant \sum_\lambda p(\lambda) \sqrt{\left(\iint_{R}\left[W_A^\lambda(x \cos \theta, p \cos \theta)\right]^2 \dd x \dd p\right)} \sqrt{\left(\iint_{R}\left[W_{f_{\theta}(\rho_B^\lambda)}^\lambda(x \sin \theta, p \sin \theta)\right]^2 \dd x \dd p\right)}  \notag\\
& \leqslant \sum_\lambda p(\lambda) \sqrt{\left(\frac{\iint_{\mathbb{R}^2}\left[W_A^\lambda(x, p)\right]^2 \dd x \dd p}{\cos ^2 \theta} \right)\left(\frac{\iint_{\mathbb{R}^2} \left[W_{f_{\theta}(\rho_B^\lambda)}^\lambda(x, p)\right]^2 \dd x \dd p}{\sin ^2 \theta}\right)}  \notag\\
& \leqslant \frac{1}{4 \pi |\cos \theta \sin \theta|} \;,
\end{align}
where $R$ is an arbitrary region in phase space and we use $W_B^\lambda\left(x^{\prime} \sin \theta, p^{\prime} \sin \theta\right)=W_{f_\theta\left(\rho_B\right)}^\lambda(x \sin \theta, p \sin \theta)$. Here, in the third-to-last row we used the Cauchy–Schwarz inequality, and in the last row we used Eq.~(\ref{eq:WignerPurityBound}). $\Box$

\vspace{2mm}
The physical motivation for criterion~\eqref{crit3} coincides with overlap calculations between two modes. For a pure product state, the fidelity between modes has a well-defined lower bound of zero. Since this expression is linear, it generalizes directly to mixed separable states. However, when the initial state is entangled, applying the same overlap formula can yield negative values because the formula no longer represents a valid fidelity between modes for entangled states. This negative value indicates non-separability.

\vspace{2mm}
\textit{Proof of criterion~\eqref{crit3}} -- Using the expression given by Eq.(\ref{eq:SeparableWigner}), we have for any separable state
\begin{align}
\iint_{-\infty}^{\infty} & W_{A B}\left(x, p, x', p'\right) \dd x \dd p \notag\\
= & \iint_{-\infty}^{\infty} \sum_\lambda p(\lambda) W_A^\lambda(x, p) W_B^\lambda\left(x', p'\right) \dd x \dd p \notag\\
= & \sum_\lambda p(\lambda) \iint_{-\infty}^{\infty} W_A^\lambda(x, p) W_{f\left(\rho_B\right)}^\lambda(x, p) \dd x \dd p \notag\\
\geqslant & 0 \;,
\end{align}
where in the last row we used Eq.(\ref{eq:WignerTraceProductBound}). $\Box$

\vspace{2mm}
To some extent, we are using a single slice of the two-mode joint Wigner function to certify entanglement. It would be beneficial to perform an optimization to find the best slice. This approach would allow us to observe the maximum violation of our criteria.
(i) To maximize criterion~\eqref{crit1}, we simultaneously optimize the transformation $(x,p)\rightarrow (x',p')$ and the parameter $\theta$ within the range $0< \theta<\pi$ and $\theta \neq \pi / 2$. The real parameters which determine the transformation Eq.~\eqref{eq:QuadratureTransformation} in Eq.~\eqref{eq:Mtransform}, must satisfy the condition $\Delta = \pm 1$.
(ii) To maximize criterion~\eqref{crit2}, we first optimize the transformation Eq.~\eqref{eq:Mtransform} and the parameter $\theta$ as previously described, across the entire plane to find the optimal correlation that maximally violates the criterion. Afterward, we focus on the region where the Wigner function is primarily distributed. In this step, we minimize the integral region $R$, while still ensuring that criterion~\eqref{crit2} remains violated. This approach allows us to further reduce the measurements needed for criterion~\eqref{crit2}, which for $R=\mathbb{R}^2$ is equivalent to performing full tomography for a single mode, thereby saving measurement resources.
(iii) To minimize criterion~\eqref{crit3}, we apply the same optimization of transformation $(x',p')$ as described previously.

Testing criterion~(\ref{crit3}) can be performed in two ways. The first is to measure it in its original form, which is equivalent to performing single-mode tomography to obtain a full slice of the two-mode Wigner function. Alternatively, if the transformation $(x',p')$ corresponds to a unitary transformation acting on $\rho_B$, a second, potentially weaker method can be applied, requiring to measure only one point of the reduced Wigner function. Specifically, let us consider a beam-splitter transformation of the modes, $\hat{a}_\pm=(\hat{a}_A \pm \hat{a}_B)/\sqrt{2}$. 
Using Eq.~\eqref{eq:BeamSplitterTransformationWithTheta} with $\theta=\pi/4$ and $\left(x^{\prime}=\sqrt{2} x_{+}-x, p^{\prime}=\sqrt{2} p_{+}-p\right)$, we arrive at the special case $\iint_{-\infty}^{\infty} W_{A B}\left(x, p, x^{\prime}, p^{\prime}\right) \dd x \dd p =\frac{1}{2} W_{+}\left(x_{+}, p_{+}\right)$, which is half the Wigner function of one of the two modes resulting from the beam-splitter transformation.
This means that, to violate criterion~(\ref{crit3}) it is enough to find a single point in the $(x_+,p_+)$ space where $W_+$ is negative~\cite{DynamicsJayachandran2023,CertifiableZaw2024}, see SM Sec.~VI~\cite{supplement}.

\vspace{2mm}
\textbf{Comparison to known criteria and applications.}--
Interestingly, we prove that Criterion \eqref{crit1} has a close connection with the entanglement negativity and can thus be used as a bound on entanglement monotones.
This originates from a connection between Criterion \eqref{crit1} and the Peres-Horodecki separability criterion~\cite{PeresPRL1996,HorokechiPLA1997}, which states that all separable states must have positive partial transpose (PPT). This follows from the fact that $\rho_{AB}^{T_B}=\sum_\lambda p(\lambda) \rho_A^\lambda \otimes (\rho_B^\lambda)^T$ is a valid state with non-negative eigenvalues. A violation of this condition indicates entanglement. Such criterion is particularly powerful in CV systems, where bound entanglement is a rare phenomenon~\cite{HorodeckiBoundEntanglement2003}. For entangled states that are not PPT, the entanglement negativity is a useful entanglement monotone defined via the negative eigenvalues of $\rho_{AB}^{T_B}$~\cite{KarolVolume1998,EisertAComparison1999,VidalComputable2002,JinhyoungLeePartial2000}, specifically, $\mathcal{N}(\rho)=(\|\rho^{T_B}\|_{\text{tr}}-1)/2$. In the SM Sec.~I~\cite{supplement}, we show that the violation $v$ of criterion~\eqref{crit1}, quantified as the difference between both sides, provides a lower bound for entanglement negativity $\mathcal{N}(\rho)\geq \pi v$. For Schmidt-correlated states, both the PPT criterion~\cite{KhasinNegativity2007} and criterion~\eqref{crit1} are sufficient and necessary conditions. Moreover, the optimal $\pi v$ can even accurately provide the exact value of the entanglement negativity.

Similarly, criterion \eqref{crit2} exhibits a fundamental connection with the computable cross norm or realignment (CCNR) criterion \cite{OliverRudolph_2000,rudolph2005further,PEREZGARCIA2004149,RudolphSome2003}, which complements the PPT criterion. 
The left-hand side of criterion \eqref{crit2} can be expressed as
\begin{align}
\text{l.h.s} =\frac{1}{2 \pi|\sin 2 \theta|} &\iint_R\Bigl| \Bigl\langle\frac{|\cos \theta|}{\sqrt{\pi}} D(x \cos \theta, p \cos \theta) \hat\Pi D^{\dagger}(x \cos \theta, p \cos \theta) \nonumber \\
&\otimes \frac{|\sin \theta|}{\sqrt{\pi}} D(x' \sin \theta,p' \sin \theta) \hat\Pi D^{\dagger}(x' \sin \theta,p' \sin \theta) \Bigr\rangle \Bigr| \dd x \dd p, \nonumber
\end{align}
where $D\hat\Pi D^\dagger$ represents the displaced parity measurement detailed in the following section. The operators $P_A(x, p)=\frac{|\cos \theta|}{\sqrt{\pi}} D(x \cos \theta, p \cos \theta) \hat\Pi D^{\dagger}(x \cos \theta, p \cos \theta)$ and $P_B(x', p')=\frac{|\sin \theta|}{\sqrt{\pi}}$ $D(x' \sin \theta,p' \sin \theta)$ $\hat\Pi$ $D^{\dagger}(x' \sin \theta,p' \sin \theta)$ are specifically chosen to be normalized such that $\operatorname{tr}(P^\dagger(x_1,p_1)P(x_2,p_2)) = \delta(x_1-x_2)\delta(p_1-p_2)$. Consequently, the integrand represents a Bloch decomposition of the quantum state \cite{AsadianHeisenberg2016}. Following the analysis presented in Ref.~\cite{YuEntanglement2005} and using properties of the trace norm, we obtain $\text{l.h.s.} \leq \frac{\|\mathcal{R}(\rho)\|_{\operatorname{tr}}}{2 \pi|\sin 2 \theta|}$, where $\mathcal{R}(\rho)$ denotes the realigned density matrix in the CCNR criterion.

In the following, we illustrate how criteria~(\ref{crit1},\ref{crit2},\ref{crit3}) are applied to several experimentally relevant scenarios. For a comparison between our entanglement criteria and nonlocality/steering criteria based on the Wigner function see SM Sec.~VII~\cite{supplement}.

\vspace{2mm}
\textit{Two-mode squeezed states.}-- 
We begin by investigating entanglement detection in two-mode squeezed thermal (TMST) states to demonstrate the applicability of our criteria to Gaussian states. Since Gaussian states have Wigner functions that remain positive across the entire phase space, criteria~(\ref{crit1},\ref{crit2}), derived from upper bound conditions, are best suited for detecting entanglement.

An arbitrary TMST state can be generated by sending one mode of a two-mode squeezed vacuum (TMSV) state through a Gaussian channel that introduces both loss and noise \cite{YuPRA2017,MartinPRD2017}. This process can be modeled by applying a quantum-limited attenuator with transmissivity $\eta$ to the TMSV state characterized by a squeezing parameter $s$, followed by a quantum-limited amplifier with gain $\cosh^2 r$.

In the simplest case of TMSV states, \ie $\eta=1,r=0$, we find that the linear transformation $(x'=x,p'=-p)$ and $\theta=\pi/4$ are optimal and allow us to detect entanglement with both criteria~(\ref{crit1},\ref{crit2}) as long as $s>0$ (See SM Sec.~VIII~\cite{supplement}). This result coincides with the conclusion obtained from Duan-Simon and PPT criteria, which are known to be necessary and sufficient~\cite{SimonPRL2000,DuanPRL,PeresPRL1996,SeparabilityHorodeckiPLA1996}.

\begin{figure}[t]
    \begin{center}
	\includegraphics[width=0.65\columnwidth]{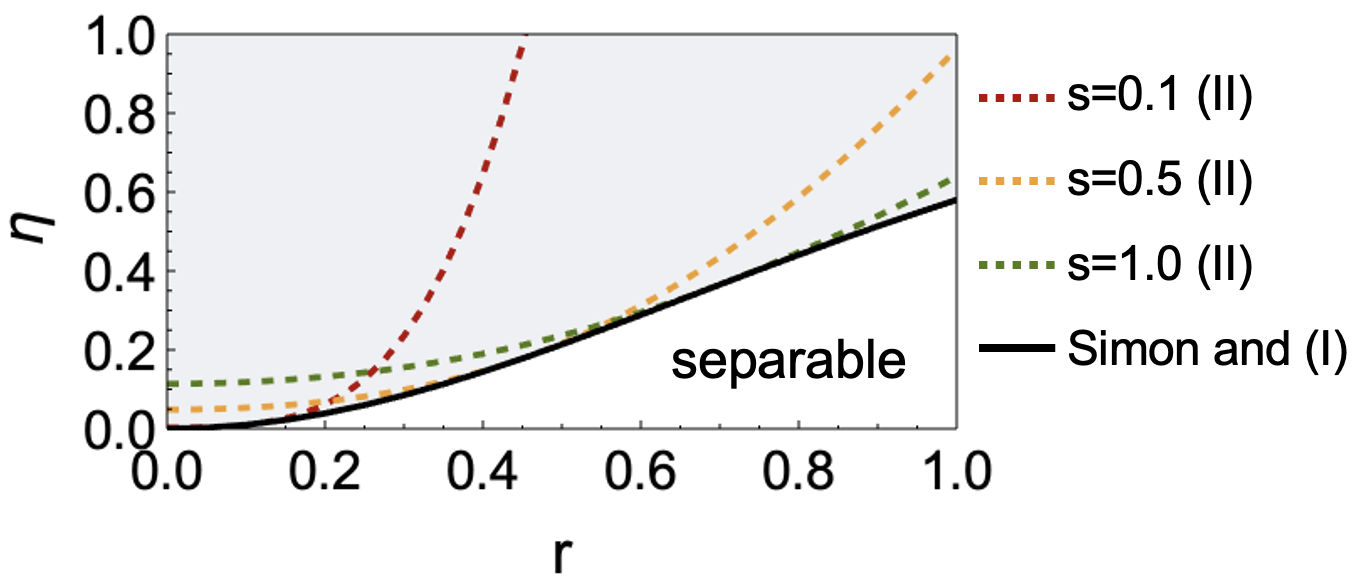}
	\end{center}
    \caption{Comparison between criteria~\eqref{crit1},~\eqref{crit2} and Simon's for TMST states. 
    Since the states considered here are Gaussian, Simon's criterion is both necessary and sufficient \cite{SimonPRL2000}. It identifies all states above the black solid line, $\eta = \tanh^2{(r)}$, as entangled (shaded region, independent of $s$).
    This bound coincides with the one given by criterion~\eqref{crit1}. On the other hand, criterion~\eqref{crit2} detects entanglement only for states above the dashed lines, each for a different $s$ value.
   }
    \label{Fig2_TMSTent}
\end{figure}

In the most general case of a TMST state, \ie $1>\eta>0,r>0$, Duan-Simon and PPT criteria are still necessary and sufficient conditions for entanglement~\cite{SimonPRL2000,DuanPRL,PeresPRL1996,SeparabilityHorodeckiPLA1996}. Remarkably, we can prove that this is also true for our criterion~\eqref{crit1}, see SM Sec.~V~\cite{supplement}. On the other hand, while our criterion~\eqref{crit2} is still useful, the bound it provides is not tight. 
In fact, for Gaussian states criterion~\eqref{crit2} is a specific case of criterion~\eqref{crit1} with $\theta=\pi/4$ and $R=\mathbb{R}^2$, and therefore \eqref{crit1} is strictly stronger than \eqref{crit2}. 
In Fig.~\ref{Fig2_TMSTent}, we illustrate this with a concrete example. The shaded region indicates TMST states that are entangled. These states are all detected by Simon's criterion, which is violated above the black solid line, since it is necessary and sufficient for Gaussian states \cite{SimonPRL2000}. The same conclusion can be obtained from our criterion~\eqref{crit1}. In addition, our criterion~\eqref{crit2} detects only a fraction of all TMST entangled states, as indicated by the regions above the dashed coloured lines. Note that these depend on the squeezing $s$, while the bound obtained from criterion~\eqref{crit1} is independent of $s$. In contrast to (\ref{crit1},\ref{crit2}), entanglement criteria in Refs.~\cite{DynamicsJayachandran2023,CertifiableZaw2024} fail to detect entanglement for any TMST states, more precisely, any Gaussian state, since they only work for states with Wigner negativity.

\vspace{2mm}
\textit{Werner states.}-- We consider here mixtures between a pure state $\ket{\psi}$ and white noise in $\{|0\rangle,|1\rangle\}^{\otimes 2}$, taking the form $\rho(\psi) = \epsilon |\psi \rangle \langle \psi |+(1-\epsilon) \mathbbm{1}/4 $, where $\epsilon\in[0,1]$ specifies the mixing weight. In the following, we will focus on two cases: $|\psi\rangle=|\Phi^+\rangle \equiv \left( |00\rangle +|11\rangle \right)/\sqrt{2}$ and $|\psi\rangle =|\Psi^+\rangle\equiv \left( |01\rangle +|10\rangle \right)/\sqrt{2}$. 

Let us start with Werner state $\rho (\Phi^+)$, whose Wigner distribution $W_{AB}$ under the linear transformation $x'=x,p'=-p$ (or equivalently, $a=1, b=c=0, d=-1, x_0=p_0=0$ in \cref{eq:QuadratureTransformation}) is non-negative for any $\epsilon$ (See SM Sec.~VIII~\cite{supplement}). Considering the integration to be over the entire space $\mathbb{R}^2$ and $\theta=\pi/4$, we obtain for criterion~\eqref{crit2} 
\begin{align}
\iint_{-\infty}^{\infty} |W_{AB} \left(x,p,x,-p\right)| \dd x \dd p =\frac{3\epsilon+1}{8\pi} \leq \frac{1}{4\pi} \;,
\end{align}
which is violated for $\epsilon>1/3$. Since $\theta=\pi / 4$ represents a balanced setting, we have moved the constant it introduces to the other side of the inequality. The same conclusion is obtained from criterion~\eqref{crit1}. This result coincides with the known tight bound for entanglement of $\rho (\Phi^+)$~\cite{SeparabilityHorodeckiPLA1996}.

For Werner state $\rho(\Psi^+)$, we find that the Wigner function $W_{AB}(x,p,-x,-p)$ is negative over the entire phase space at $\epsilon=1$ (See SM Sec.~VIII~\cite{supplement}). We thus calculate criterion~\eqref{crit3} and obtain
\begin{align}
\iint_{-\infty}^{\infty} W_{AB} \left(x,p,-x,-p\right) \dd x \dd p =\frac{1-3\epsilon}{8\pi} \geq 0 \;,
\end{align}
which is violated for $\epsilon>1/3$. Also this result coincides with the known tight bound for entanglement of $\rho (\Psi^+)$~\cite{SeparabilityHorodeckiPLA1996}.

In summary, we have shown that our criteria based on the Wigner function provide bounds that are tight for the Werner states considered, since they coincide with the ones obtained from PPT.

\begin{figure}[t]
    \begin{center}
	\includegraphics[width=85mm]{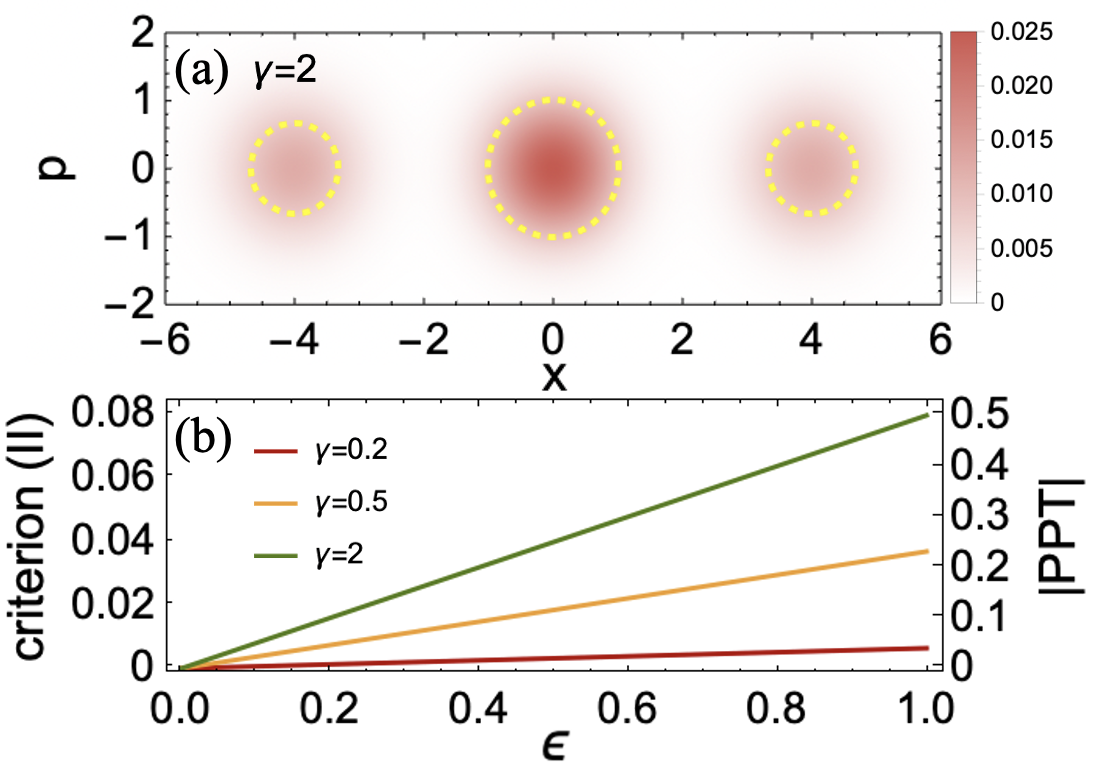}
	\end{center}
    \caption{Entanglement detection for dephased cat states Eq.~\eqref{eq:rhocat}. (a) Slice of the joint Wigner function defined by $W_{AB}(x,p,x,-p)$, for $\gamma=2$ and $\epsilon=1$. Yellow dashed circles indicate the smallest region $R$ needed to violate criterion~\eqref{crit2}. (b) Amount of violation for criterion~\eqref{crit2} with $\theta=\pi/4$, i.e. $\iint_{-\infty}^{\infty}|W_{AB}(x,p,x,-p)|\dd x \dd p-1/(4\pi)$, and for PPT criterion, i.e. $|\lambda_{\min}(\rho^{T_B})|$, as a function of $\epsilon$ and for three values of $\gamma$. Entanglement can be revealed by both criteria whenever $\gamma,\epsilon>0$. }
    \label{Fig3_Catent}
\end{figure}

\vspace{2mm}
\textit{Dephased two-mode cat states.}--
To illustrate the performance of our criteria on non-Gaussian states with many excitations, we consider states of the form
\begin{align}\label{eq:rhocat}
\rho=\epsilon|\psi\rangle \langle \psi |+\frac{1-\epsilon}{2}\left(|\gamma,\gamma\rangle \langle \gamma,\gamma|+|-\gamma,-\gamma\rangle \langle -\gamma,-\gamma| \right) \;,
\end{align}
where $|\psi\rangle=\mathcal{N}(|\gamma,\gamma\rangle+|-\gamma,-\gamma\rangle)$ is a two-mode superposition of coherent states $|\gamma\rangle$, i.e. a two-mode cat state, with $\mathcal{N}$ a normalization constant, and $\epsilon\in[0,1]$ determines the mixing weight in terms of a dephasing of the superposition.

As illustrated in Fig.~\ref{Fig3_Catent}(a), the joint Wigner distribution under the transformation $(x'=x,p'=-p)$ and $\theta=\pi/4$, which we numerically find to be optimal, is positive over the entire space. When computing criteria~(\ref{crit1},\ref{crit2}) and PPT criterion we find that they are violated for any $\gamma, \epsilon>0$ (See SM Sec.~VIII~\cite{supplement}). The value for the criteria is shown in Fig.~\ref{Fig3_Catent}(b). 
For the cat states considered, our criteria~(\ref{crit1},\ref{crit2}) are more powerful in detecting entanglement than the entanglement criterion discussed in Refs.~\cite{DynamicsJayachandran2023,CertifiableZaw2024}, since the latter has the same entanglement bound as criterion~\eqref{crit3}, which is useful but not tight for $\rho(\psi^+)$ (See SM Sec.~VIII~\cite{supplement}).
Also in practical considerations, different from PPT criterion and entanglement criterion based on quantum Fisher information \cite{ManuelPRA2016} where full information of the density matrix is required, the criteria~(\ref{crit1},\ref{crit2}) only need partial information (a slice of Wigner function) determined by the transformation $\left(x',p'\right)$. Additionally, for criterion~\eqref{crit2} the required integral region can be further narrowed down to a limited region, for example, the yellow circles in Fig.~\ref{Fig3_Catent}(a).

To bridge the gap between our theoretical criteria and practical implementations with current technology, we provide a comprehensive error analysis from two perspectives in SM Sec. IX~\cite{supplement}. First, we examine the impact of grid spacing when using discrete-point measurements. When implementing our criteria experimentally, the continuous Wigner function must be sampled on a discrete grid through displaced parity measurements. We demonstrate that this discretization introduces minimal errors--even for coarse sampling with grid spacing $\delta = 0.5$--confirming the robustness of our approach. Second, we present a measurement strategy specifically designed for our criteria. We introduce a measurement protocol that evaluates our criteria through scanning measurements uniformly distributed over the relevant phase-space region (for example, using Fermat spiral patterns \cite{VogelSunflower1979} or randomized measurement). This approach is statistically equivalent to single-point displaced parity measurements, thereby significantly reducing data requirements. For the paradigmatic two-mode squeezed vacuum state, our method requires two orders of magnitude fewer measurements than the four-point Bell inequality test \cite{BanaszekPRL1999,BanaszekPRA1998} to achieve comparable confidence levels. We further illustrate in SM Sec. IX~\cite{supplement} that, when the total number of measurements is expressed as $N_{\mathrm{tot}}=M \cdot N_{\mathrm{shot}}$, where $M$ denotes the number of measurement rounds and $N_{\mathrm{shot}}$ the number of shots per round, the randomized protocol requires only $M=1$ for criteria~\eqref{crit1} and~\eqref{crit3}, and $M=2$ for criterion~\eqref{crit2}. This represents a significant reduction compared to full two-mode tomography, where $M$ typically scales as $D^4$ with $D$ being the effective truncation dimension of each mode.

\begin{figure}[t]
    \begin{center}
	\includegraphics[width=\textwidth]{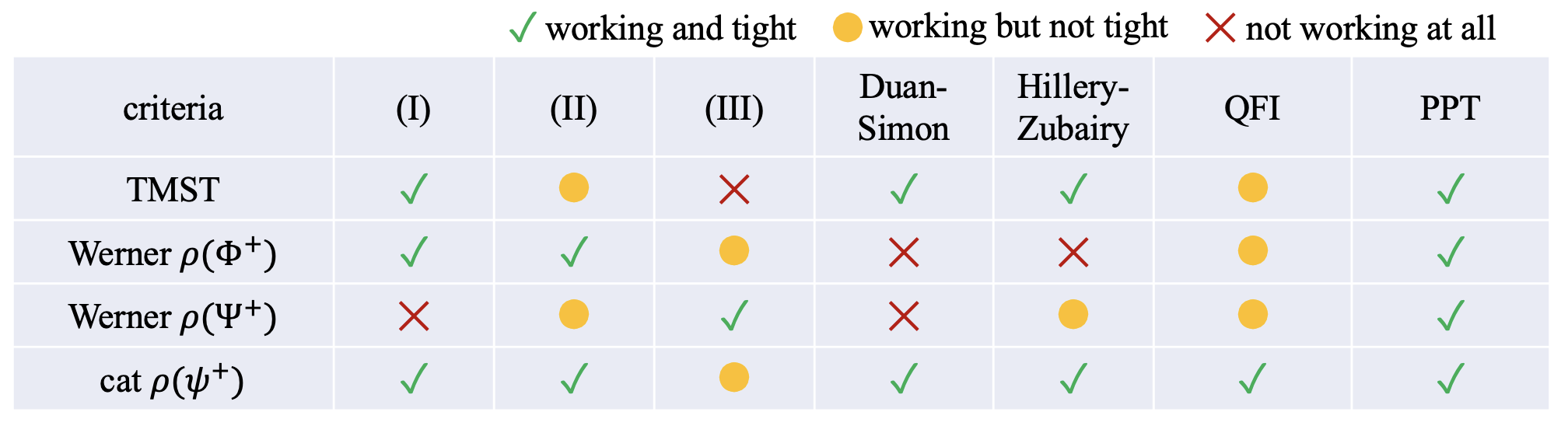}
	\end{center}
    \caption{Comparison of the entanglement detection capabilities of different criteria, \ie the criteria~(\ref{crit1},\ref{crit2},\ref{crit3}), Duan-Simon criterion~\cite{SimonPRL2000,DuanPRL}, Hillery-Zubairy criterion~\cite{HilleryPRL2006}, QFI criterion~\cite{ManuelPRA2016} and PPT criterion~\cite{PeresPRL1996,SeparabilityHorodeckiPLA1996}, for TMST, Werner and dephased cat states.}
    \label{Fig:SMtable}
\end{figure}

\vspace{2mm}
\textbf{Summary and outlook.}-- We have introduced criteria to detect CV bipartite entanglement from the Wigner function. By comparing them with known criteria from the literature \cite{SimonPRL2000,DuanPRL}, we have shown cases where they are necessary and sufficient. 
Crucially, our criteria require only partial information on the system's state and can be implemented experimentally by direct Wigner function measurements. 
Importantly, the amount of information required is no more than a single-mode Wigner tomography.

Beyond the specific examples demonstrated, each criterion targets distinct classes of entangled states. Criteria \eqref{crit1} and \eqref{crit2} provide lower bounds on the entanglement negativity and the degree of CCNR criterion violation, respectively. Criterion \eqref{crit3}, in the meanwhile, specifically detects entangled states that exhibit Wigner negativity.
For a detailed comparison, see Fig.~\ref{Fig:SMtable}.

It would be interesting to investigate whether criteria similar to~(\ref{crit1},\ref{crit2},\ref{crit3}) can be found for discrete variable systems, using the Wigner function for spin states \cite{DowlingPRA94}. This would allow for alternative methods for the detection of entanglement between \eg atomic ensembles \cite{YumangNJP2019,MatteoPRA2020,Kitzinger21,jiajiePRAcondSS,fadel_multiparameter_2023}. We note that our approach readily extends beyond beamsplitters to encompass more general two-mode transformations of phase-space coordinates, including operations such as two-mode squeezing.

\vspace{2mm}
\textbf{Acknowledgments.}-- We are indebted to M. Walschaers for inspiring and insightful discussions on the topic and to L. H. Zaw and E. Agudelo for valuable comments on our work.
This work is supported by the National Natural Science Foundation of China (Grants No. 12125402, No. 12534016, No. 12405005, No. 12505010, No. 12447157), the Beijing Natural Science Foundation (Grant No. Z240007), and the China Postdoctoral Science Foundation (No. 2023M740119 and 2024M760072). J. Guo acknowledges the CPSF Postdoctoral Fellowship Program (No. GZB20240027).
MF was supported by the Swiss National Science Foundation Ambizione Grant No. 208886, and by the Branco Weiss Fellowship -- Society in Science, administered by the ETH Z\"{u}rich.

\bibliographystyle{quantum}
\bibliography{Reference}

\clearpage
\newpage

\section*{Supplemental material for ``Quantum entanglement in phase space''}

\section*{I. Criterion~(\ref{crit1}) can lower bound the entanglement negativity}

If we consider a beam-splitter transformation $S$ and a time reversal $(x,p)\rightarrow (x,-p)$, the left-hand side of Criterion~\eqref{crit1} can be re-expressed as
\begin{align}
\iint_{-\infty}^{\infty} \dd x \dd p W(\cos{\theta} x, \cos{\theta} p, \sin{\theta}x', \sin{\theta}p') &=   \iint_{-\infty}^{\infty} \dd x \dd p W(\cos{\theta} x, \cos{\theta} p, \sin{\theta}x, -\sin{\theta}p) \nonumber \\
&= \iint_{-\infty}^{\infty} \dd x \dd p W_{\rho^{T_B}}(\cos{\theta} x, \cos{\theta} p, \sin{\theta}x, \sin{\theta}p) \nonumber \\
&= \iint_{-\infty}^{\infty} \dd x \dd p W_{S(\rho^T_B)} (x,p,0,0) \nonumber \\
&= W_{\text{tr}_{A'}[S(\rho^{T_B})] } (0,0).
\end{align}
Such Wigner function can be expressed by the displaced parity operators $\hat{P}(\alpha)=\hat{D}^\dagger(\alpha) (-1)^{\hat{n}} \hat{D}(\alpha)$ 
\begin{align}
2\pi W_{\text{tr}_{A'}[S(\rho^{T_B})] } (0,0) &=  \text{tr}\left[ \hat{P}_{B'}(0) \text{tr}_{A'}[S(\rho^{T_B})] \right] \nonumber \\
&=  \text{tr}\left[ (\hat{I}_{A'} \otimes \hat{P}_{B'}(0))\cdot S (\rho^{T_B}) \right].
\end{align}
Here, $(\hat{I}_{A'}\otimes \hat{P}_{B'})$ and the beam-splitter transformation S can be regarded as a unitary operator $\hat{U}^\prime$. We can therefore obtain
\begin{align}
\text{tr}\left[ \hat{U}^\prime \rho^{T_B} \right] \leq \max_{\hat{U}} \text{tr}[\hat{U}\rho^{T_B}] =||\rho^{T_B}||_{\text{tr}}.
\end{align}
Thus, We conclude that Criterion~\eqref{crit1} provides a lower bound of the entanglement negativity $\mathcal{N}$
\begin{align}
\mathcal{N}(\rho) &\equiv \frac{1}{2} \left( ||\rho^{T_B}||_{\text{tr}}-1 \right)  \nonumber \\
& \geq \pi \iint_{-\infty}^{\infty} \dd x \dd p W(\cos{\theta} x, \cos{\theta} p, \sin{\theta}x, -\sin{\theta}p) -\frac{1}{2}
\end{align}

In fact, the bound is tight for Schmidt-correlated states $\sigma = \sum_{nm} c_{nm} |n,n\rangle \langle m,m |$, where $c_{nm}$ are nonnegative real coefficients. We will prove it in the following.

The partial transpose on mode B yields $\sigma^{T_B} = \sum_{nm} c_{nm} |n,m\rangle \langle m,n|$. The norm of $\sigma^{T_B}$ is $||\sigma^{T_B} ||  \equiv  \sqrt{ \left( \sigma^{T_B} \right)^\dagger \left( \sigma^{T_B} \right) } =   \sqrt{ \sum_{n,m}  c^2_{nm}  |m,n\rangle  \langle m,n|   } =  \sum_{n,m}  c_{nm}  |m,n\rangle  \langle m,n|$. By multiplying a swap operator $S_{\text{swap}}=\sum_{n,m} |m,n\rangle \langle n,m|$, we have $\sigma^{T_B} S_{\text{swap}} = \sum_{n',m',n,m}  c_{nm}  (|n,m\rangle  \langle m,n|)$ $(|m',n'\rangle \langle n',m'|) = \sum_{n,m}  c_{nm}  |m,n\rangle  \langle m,n|$. Thus it is found that $||\sigma^{T_B}|| = \sigma^{T_B} S_{\text{swap}} $.

Therefore, we can further obtain 
\begin{align}
||\sigma^{T_B} ||_{\text{tr}} &= \text{tr}\left[ \sigma^{T_B} S_{\text{swap}} \right] \nonumber \\
&= \iint_{-\infty}^{\infty} \dd x_1 \dd p_1 \dd x_2 \dd p_2 W_{\sigma^{T_B}}(x_1, p_1,  x_2, p_2) \cdot 4\pi \delta \left( x_1-x_2 \right) \delta \left( p_1-p_2 \right) \nonumber \\
&= 4\pi \iint_{-\infty}^{\infty} \dd x \dd p W_{\sigma^{T_B}}\left(x, p,  x, p \right) \nonumber \\
&= 4\pi \iint_{-\infty}^{\infty} \dd x \dd p W_{\sigma}\left(x, p,  x, -p \right) \nonumber \\
&= 2\pi \iint \dd x \dd p W_{\sigma}(x/\sqrt{2}, p/\sqrt{2}, x/\sqrt{2}, -p/\sqrt{2} ).
\end{align}
Here, we use the Wigner function of a swap operator $4\pi \delta \left( x_1-x_2 \right) \delta \left( p_1-p_2\right)$. The last row is the left-hand side of Criterion~\eqref{crit1} under the transformation with a time-reversal $(x,p)\rightarrow(x,-p)$ and $\theta=\pi/4$.

\section*{II. Choice of $\Delta=\pm 1$ in linear transformations}

\subsection*{A. Criterion~\eqref{crit1} requires $\Delta=-1$}

Criterion~\eqref{crit1} can detect entanglement if and only if $\Delta=-1$. Now we prove it by demonstrating that Criterion~\eqref{crit1} will not be violated for any states if $\Delta=+1$. 

In the case of $\Delta=+1$, the quadratures ($x',p'$) are obtained from ($x,p$) through a unitary transformation, i.e. $x'=U^\dagger x U, p'=U^\dagger p U$. 
Then according to Eq.~\eqref{eq:WCosSinThetaReduceUpperBound} and Eq.~\eqref{eq:CriterionIproof} of Sec.~IV in SM, Criterion~\eqref{crit1} can be expressed as
\begin{align}
& \iint_{-\infty}^{\infty} W_{AB} (\cos{\theta} x, \cos{\theta} p, \sin{\theta} x', \sin{\theta} p') \dd x \dd p \nonumber \\
& = \iint_{-\infty}^{\infty} W_{Ah_2h_1(B)} \left( \cos{\theta}x+\sin{\theta}X, \cos{\theta} p+\sin{\theta}P,\sin{\theta}x-\cos{\theta}X,\sin{\theta}p-\cos{\theta}P \right) \dd x \dd p \nonumber \\
& = W_{B'} (X,P) \nonumber,
\end{align}
where $(X,P)$ are constant coordinates, and $B'$ is an output mode of the beam-splitter with input being $\rho'_{AB}=\mathbbm{1} \otimes h_2 h_1 (\rho_{AB})$.
For any initial states $\rho_{AB}$, the state $\rho'_{AB}$ after the unitary transformation is 
\begin{align}
\rho'_{AB}= \left( \mathbbm{1} \otimes U \right) \rho_{AB} \left( \mathbbm{1} \otimes U \right)^\dagger =U^{'} \rho_{AB} U^{'\dagger},
\end{align}
where $h_1, h_2$ are state transformations in Eqs.~(\ref{eq:h1},\ref{eq:h2}) and $U,U'$ are unitary transformation on mode B and modes AB, respectively. After the unitary transformation, $\rho'_{AB}$ should be a physical state, so that $B'$ is also physical.  Therefore, the Wigner function of the reduced mode $B':(X,P)$ is always upper-bounded by $1/(2\pi)$,  meaning Criterion~\eqref{crit1} will not be violated.

\subsection*{B. Criterion~\eqref{crit3} requires $\Delta=+1$}

Criterion~\eqref{crit3} can detect entanglement if and only if $\Delta=+1$. We prove that Criterion~\eqref{crit3} holds for all states if $\Delta=-1$.

$\Delta=-1$ indicates a mirror reflection, as an example we consider the time reversal $(x,p) \rightarrow (x,-p)$. The local linear coordinate transformation corresponds to partial transpose of the state $W_{\rho} (x_A,p_A,x_B,-p_B) = W_{\rho^{T_B}} (x_A,p_A,x_B,p_B)$. 
Define
\begin{align}
I_0 := \iint_{-\infty}^{\infty} W_\rho(x, p,\, x, -p)\, \mathrm{d} x\, \mathrm{d} p.
\end{align}
Substituting
\begin{align}
W(\boldsymbol{\xi}) = \int \frac{\mathrm{d}^{4}\boldsymbol{\alpha}}{(2\pi)^{4}}\, \exp\left(-i\, \boldsymbol{\xi}^{\mathsf{T}} \Omega\, \boldsymbol{\alpha}\right) \chi(\boldsymbol{\alpha}),
\end{align}
we obtain
\begin{align}
I_0 = \int \frac{\mathrm{d}^{4}\boldsymbol{\alpha}}{(2\pi)^{4}}\, \chi(\boldsymbol{\alpha}) \iint \mathrm{d} x\, \mathrm{d} p\; \exp\Big(-i\big[x(\alpha_{p_A}+\alpha_{p_B}) - p(\alpha_{x_A}-\alpha_{x_B})\big]\Big).
\end{align}
The integrals over $x$ and $p$ yield Dirac delta functions, $\int \mathrm{d} x\, e^{-ix(\alpha_{p_A}+\alpha_{p_B})} = 2\pi\,\delta(\alpha_{p_A}+\alpha_{p_B})$,\\ $\int \mathrm{d} p\, e^{ip(\alpha_{x_A}-\alpha_{x_B})} = 2\pi\,\delta(\alpha_{x_A}-\alpha_{x_B})$, so that
\begin{align}\label{eq:I0char}
I_0 = \frac{1}{(2\pi)^{2}} \iint \mathrm{d} \alpha_x\, \mathrm{d} \alpha_p\; \chi(\alpha_x, \alpha_p,\, \alpha_x, -\alpha_p).
\end{align}
Using $\chi(\boldsymbol{\alpha}) = \operatorname{Tr}\left[\rho\, e^{i\hat{\boldsymbol{\xi}}^{\mathsf{T}}\Omega\,\boldsymbol{\alpha}}\right]$, we can obtain $I_0 = \operatorname{Tr}[\rho\, K]$, where
\begin{align}
K := \frac{1}{(2\pi)^{2}} \iint \mathrm{d} \alpha_x\, \mathrm{d} \alpha_p\; \exp\Big(i\big[\alpha_p(\hat{x}_A - \hat{x}_B) - \alpha_x(\hat{p}_A + \hat{p}_B)\big]\Big).
\end{align}
Define $\hat{X}_{-} := \hat{x}_A - \hat{x}_B$ and $\hat{P}_{+} := \hat{p}_A + \hat{p}_B$. Since $\hat{X}_{-}$ and $\hat{P}_{+}$ commute, the exponential factorizes as $e^{i(\alpha_p \hat{X}_{-} - \alpha_x \hat{P}_{+})} = e^{i\alpha_p \hat{X}_{-}}\, e^{-i\alpha_x \hat{P}_{+}}$, and
\begin{align}
K = \delta(\hat{X}_{-})\,\delta(\hat{P}_{+}) = \delta(\hat{x}_A - \hat{x}_B)\,\delta(\hat{p}_A + \hat{p}_B).
\end{align}
Since $K \geq 0$, it follows that
\begin{align}
I_0 = \operatorname{Tr}[\rho\, K] \geq 0 .
\end{align}
For a general $\Delta = -1$ transformation $M = S R$ where $S$ is symplectic and $R$ is the standard reflection, the corresponding operator is $K_M$ is unitarily equivalent to $K$. An additional displacement can be considered in a similar way. So the conclusion extends to all $\Delta = -1$ cases. Therefore, Criterion~\eqref{crit3} holds for all states when $\Delta = -1$ and cannot be used to detect entanglement.

\section*{III. An intuitive phase-space counterpart to the Duan-Simon criterion and their comparison}

Using Williamson’s theorem, the covariance matrix of a single-mode Gaussian state can be diagonalized as $\sigma=\operatorname{diag}(l,l)$ through a symplectic transformation, where $l\geqslant 1$ due to the uncertainty relation. 
Conversely, if $0<l<1$ the purity $1 / \sqrt{\operatorname{Det} \sigma}$ would exceed $1$.
Consequently, from the purity of the partially transposed state we can establish an entanglement criterion based on the phase space distribution.

\textbf{Observation S1.} Consider first performing a transposition on mode B, followed by applying an arbitrary beam-splitter operation $S_{\mathrm{BS}}=\left(\begin{smallmatrix}\cos (\theta) & \sin (\theta) \\ \sin (\theta) & -\cos (\theta)\end{smallmatrix}\right)$ to the two modes, with the splitting ratio determined by the angle $\theta$. The coordinate transformations are given by $\left(x_A^{\prime}, x_B^{\prime}\right)^T=S_{\mathrm{BS}}\left(x_A, x_B\right)^T$, $\left(p_A^{\prime}, p_B^{\prime}\right)^T=S_{\mathrm{BS}}\left(p_A,-p_B\right)^T$. For any separable bipartite state $\rho_{AB}$ we have 
\begin{equation}\label{eq:critReducedPurity}
4 \pi \iint_{-\infty}^{\infty}\left[W^{\prime}\left(x_B^{\prime}, p_B^{\prime}\right)\right]^2 \dd x_B^{\prime} \dd p_B^{\prime} \leqslant 1,
\end{equation}
where $W_{B^{\prime}}\left(x_B^{\prime}, p_B^{\prime}\right)=\iint \mathrm{d} x_A^{\prime} \mathrm{d} p_A^{\prime} W_{A B}^{\left(\mathrm{BS} \circ T_B\right)}\left(x_A^{\prime}, p_A^{\prime}, x_B^{\prime}, p_B^{\prime}\right)$ represents the state of one output mode after the beam-splitter transformation. Note that a transposition on mode B is not a physical operation, so that the measurements required are experimentally challenging and effectively equivalent to a full state tomography.

\vspace{2mm}
\textbf{Proof.} If $\rho_{AB}$ is separable, then its partial transpose $\rho_{AB}^{T_B}$ remains a physical state \cite{SimonPRL2000}. Applying a beam-splitter transformation to $\rho_{AB}^{T_B}$ must also result in a physical state with its purity being upper bounded by $1$. 
\qed

In the following we show that this criterion is equivalent to the Duan-Simon entanglement criterion for the CV EPR states commonly realized experimentally. The standard form of a Gaussian state $\rho_G$ is given by~\cite{DuanPRL}
\begin{equation}\label{eq:CMStandardForm}
V=\left(\begin{array}{cccc}
n & 0 & c_1 & 0 \\
0 & n & 0 & c_2 \\
c_1 & 0 & m & 0 \\
0 & c_2 & 0 & m
\end{array}\right),
\end{equation}
In the following, we consider the case $c_1=-c_2=c$, which is often satisfied by commonly experimentally realized CV EPR states, such as two-mode squeezed thermal states. According to Simon's criterion, all separable states satisfy~\cite{SimonPRL2000}
\begin{equation}\label{eq:SimonCriterion}
V+i \tilde{\Omega} \geqslant 0, \quad \tilde{\Omega}=\left(\begin{array}{cccc}
0 & 1 & 0 & 0 \\
-1 & 0 & 0 & 0 \\
0 & 0 & 0 & -1 \\
0 & 0 & 1 & 0
\end{array}\right).
\end{equation}
The sufficient and necessary condition for $\rho_G$ to be entangled can thus be expressed as
\begin{equation}
c^2>(n-1)(m-1).
\end{equation}
Next, we demonstrate that this inequality can also be derived from our observation. The Wigner function of $\rho_G$ is given by
\begin{equation}
W_{A B}\left(x_A, p_A, x_B, p_B\right)=\frac{1}{4 \pi^2\left(m n-c^2\right)} \exp \left(\frac{2 c p_B p_A-2 c x_A x_B+m p_A^2+m x_A^2+n p_B^2+n x_B^2}{2 c^2-2 m n}\right).
\end{equation}
By performing a partial transposition on mode $B$ and then sending the two modes through a beam-splitter $S_{\mathrm{BS}}=\left(\begin{smallmatrix}\cos (\theta) & \sin (\theta) \\ \sin (\theta) & -\cos (\theta)\end{smallmatrix}\right)$, which corresponds to the coordinate transformations $\left(x_A^{\prime}, x_B^{\prime}\right)^T=S_{\mathrm{BS}}\left(x_A, x_B\right)^T$, $\left(p_A^{\prime}, p_B^{\prime}\right)^T=S_{\mathrm{BS}}\left(p_A,-p_B\right)^T$, we obtain the Wigner function of one output mode of the beam-splitter,
\begin{align}
W_B^{\prime}(x_B^{\prime},p_B^{\prime}) &=\frac{1}{2 \pi\left(-2 c \sin (\theta) \cos (\theta)+m \cos ^2(\theta)+n \sin ^2(\theta)\right)} \nonumber \\
& \times \exp \left(-\frac{x_B^{\prime 2}+p_B^{\prime 2}}{2\left(-2 c \sin (\theta) \cos (\theta)+m \cos ^2(\theta)+n \sin ^2(\theta)\right)}\right).
\end{align}
The purity of this output mode is given by
\begin{equation}
\operatorname{Tr}\left[\left(\rho_B^{\prime}\right)^2\right]=\frac{2}{m+n-2 c \sin (2 \theta)-\cos (2 \theta)(n-m)}.
\end{equation}
Without losing generality, assume $n>m$. The purity reaches its maximum value of $2/(m+n-\sqrt{4 c^2+(n-m)^2})$ when choosing $\tan (2 \theta)=2c/(n-m)$. The value exceeds $1$ when $c^2>(n-1)(m-1)$, indicating that $\rho_G$ is entangled. Thus, the conditions derived from both methods are equivalent.

\section*{IV. Detailed proof of criterion~(\ref{crit1})}

Let us first consider an arbitrary coordinate transformation $(x',p')$ defined as
\begin{equation}
\left(\begin{array}{c}
x' \\
p'
\end{array}\right)=\left(\begin{array}{cc}
a & b \\
c & d
\end{array}\right)\left(\begin{array}{c}
x \\
p
\end{array}\right)+
\left(\begin{array}{c}
x_0  \\
p_0 
\end{array}\right).
\end{equation}
Let $0\leqslant\theta<\pi$ be a constant angle ($\theta \neq \pi / 2$), and let $(X,P)$ represent a constant coordinate. Then, define the following two coordinate transformations according to $(x',p')$, along with the corresponding state transformations $h_1,h_2$,
\begin{equation} \label{eq:h1}
h_1: \left( \begin{array}{c}
  x\\
  p
\end{array} \right) \rightarrow \left( \begin{array}{cc}
  a & b\\
  c & d
\end{array} \right) \left( \begin{array}{c}
  x\\
  p
\end{array} \right) + \left( \begin{array}{c}
  \sin \theta x_0\\
  \sin \theta p_0
\end{array} \right) \;,
\end{equation}
\begin{equation} \label{eq:h2}
h_2: \left( \begin{array}{c}
  x\\
  p
\end{array} \right) \rightarrow \left( \begin{array}{cc}
  1 & 0\\
  0 & 1
\end{array} \right) \left( \begin{array}{c}
  x\\
  p
\end{array} \right) + \left( \begin{array}{c}
  \frac{X}{\cos \theta}\\
  \frac{P}{\cos \theta}
\end{array} \right) \;.
\end{equation}
We obtain the following expression through the coordinate transformations,
\begin{equation}\label{eq:WCosSinThetaReduceUpperBound}
\begin{aligned}
& \iint_{-\infty}^{\infty} W_{A B}\left(\cos \theta x, \cos \theta p, \sin \theta x', \sin \theta p'\right) \dd x \dd p \\
= & \iint_{-\infty}^{\infty} W_{A h_1\left(B\right)}(\cos \theta x, \cos \theta p, \sin \theta x, \sin \theta p) \dd x \dd p \\
= & \iint_{-\infty}^{\infty} W_{A h_2 h_1\left(B\right)}\left(\cos \theta x, \cos \theta p, \sin \theta x-\frac{X}{\cos \theta}, \sin \theta p-\frac{P}{\cos \theta}\right) \dd x \dd p \\
= & \iint_{-\infty}^{\infty} W_{A h_2 h_1\left(B\right)}\left(\cos \theta x+\sin \theta X, \cos \theta p+\sin \theta P, \sin \theta x-\cos \theta X, \sin \theta p-\cos \theta P\right) \dd x \dd p
\end{aligned}
\end{equation}
where in the last row we have applied the substitution $\left(x \rightarrow x+\frac{\sin \theta}{\cos \theta} X, p \rightarrow p+\frac{\sin \theta}{\cos \theta} P\right)$. Now, assume that $\rho_{AB}$ is a separable pure state, meaning $\mathbbm{1}\otimes h_2 h_1(\rho_{AB})$ should also correspond to a physical state. Next, perform a beam-splitter transformation given by $\left(\begin{smallmatrix}\cos \theta & \sin \theta \\ \sin \theta & -\cos \theta\end{smallmatrix}\right)$ on the two modes of $\mathbbm{1}\otimes h_2 h_1(\rho_{AB})$. The resulting Wigner function is given by \\
$W_{A h_2 h_1\left(B\right)}\left(\cos \theta x+\sin \theta X, \cos \theta p+\sin \theta P, \sin \theta x-\cos \theta X, \sin \theta p-\cos \theta P\right)$ where 
$A':\left(x,p\right)$ and $B':\left(X,P\right)$ represent the two output modes. Therefore, Eq.~(\ref{eq:WCosSinThetaReduceUpperBound}) becomes the Wigner function of a reduced mode, which is upper-bounded by $1/(2 \pi)$, 
\begin{equation} \label{eq:CriterionIproof}
\begin{aligned}
& \iint_{-\infty}^{\infty} W_{A h_2 h_1\left(B\right)}\left(\cos \theta x+\sin \theta X, \cos \theta p+\sin \theta P, \sin \theta x-\cos \theta X, \sin \theta p-\cos \theta P\right) \dd x \dd p \\
& =W_{B^{\prime}}\left(X, P\right) \\
& \leqslant \frac{1}{2 \pi}.
\end{aligned}
\end{equation}
Any mixed separable state must satisfy this inequality since it holds for each pure state component.

\section*{V. Criterion~(\ref{crit1}) is sufficient and necessary for Gaussian states}

The covariance matrix of any Gaussian state can be transformed into a standard form Eq.\eqref{eq:CMStandardForm} using local unitary operations~\cite{DuanPRL}. The state is entangled if and only if it violates Eq.\eqref{eq:SimonCriterion}~\cite{SimonPRL2000}. This entanglement condition can be rewritten as
\begin{equation}\label{eq:GaussianEntanglementConditionNMC1C2}
\left(\left|c_1 c_2\right|-1\right)^2<\left(c_1^2+c_2^2\right) m n+m^2+n^2-m^2 n^2.
\end{equation}
Then we show that it can also be obtained from our criterion. Select $(x',p')=\left(t x, -\frac{p}{t}\right)$. We obtain
\begin{equation}
\begin{aligned}
& W_{AB} (\cos \theta x, \cos \theta p, \sin \theta x', \sin \theta p') = \frac{1}{4 \pi^2  \sqrt{\left( c_1^2 - mn\right)  \left( c_2^2 - mn \right)}} \times \\
&\exp \left\{ \left[m t^2 \cos^2 (\theta) \left( - c_1^2 p^2 - c_2^2 x^2 + mn (p^2 + x^2) \right) \right.\right.\\
&\left.\left. + \sin (\theta) \left( n \sin(\theta) \left( - c_1^2 p^2 - c_2^2 t^4 x^2 + mn (p^2 + t^4 x^2)\right) + 2 c_2 t \cos (\theta) \left( c_1 c_2 t^2 x^2 + mnp^2 \right) \right) \right.\right.\\
&\left.\left. - c_1 t \sin (2 \theta) \left( c_1 c_2 p^2 + m n t^2 x^2 \right)\right]/\left(2 t^2  \left( c_1^2 - mn \right)  \left( m n - c_2^2 \right)\right) \right\}
\end{aligned}
\end{equation}
and its integral
\begin{equation}
\begin{aligned}
I&=\iint_{- \infty}^{\infty} W_{AB} (\cos \theta x, \cos \theta p, \sin \theta x', \sin \theta p') \dd x \dd p \\
&= \frac{1}{2 \pi \sqrt{\frac{\left( m + n t^2 + \cos (2 \theta) (m - n t^2) - 2 c_1 t \sin (2 \theta) \right) \left( m t^2 + n + \cos (2 \theta) (m t^2 - n) + 2 c_2 t \sin (2 \theta) \right)}{4 t^2}}}.
\end{aligned}
\end{equation}
The value of $I$ reaches its maximum
\begin{equation}
I = \frac{1}{2 \pi \sqrt{\frac{1}{2}  \left( - \sqrt{4 mn \left(
c_1^2 + c_2^2 \right) - 2 n^2  \left( m^2 - 2 | c_1 c_2 | \right) + 4 | c_1 c_2 | m^2 + m^4 + n^4} + 2 | c_1 c_2 | +
m^2 + n^2 \right)}}
\end{equation}
when 
\begin{equation}
\begin{aligned}
t^2 & =\frac{\left|c_2\right| m+\left|c_1\right| n}{\left|c_1\right| m+\left|c_2\right| n}, \\
\tan (2 \theta) & =\frac{-2 \sqrt{\left|c_2\right| m+\left|c_1\right| n} \sqrt{\left|c_1\right| m+\left|c_2\right| n}}{(m-n)(m+n)}.
\end{aligned}
\end{equation}
By simplifying the entanglement condition $I>1/(2\pi)$, we arrive precisely at Eq.~(\ref{eq:GaussianEntanglementConditionNMC1C2}).
Note that when $m=n$, the optimum is reached in the limit $\theta\to\pi/4$, and when $c_1=c_2=0$ the state is separable and $I\leq1/(2\pi)$. Both cases remain consistent with Eq.~(\ref{eq:GaussianEntanglementConditionNMC1C2}).

\section*{VI. A comparison between criterion~(\ref{crit3}) and previous criteria based on Wigner functions}

Reference~\cite{DynamicsJayachandran2023} introduces an entanglement criterion based on reduced states. The Wigner function is first expressed in the $\left\{\hat{a}_{+}, \hat{a}_{-}\right\}$ basis as $W_\rho\left(\alpha_{+}, \alpha_{-}\right)$. By tracing out the $a_-$ mode, the reduced Wigner function is obtained as $W_{\text{tr}_{-}(\rho)}\left(\alpha_{+}\right)=\int \dd^2 \alpha_{-} W_\rho\left(\alpha_{+}, \alpha_{-}\right)$. If any point in the reduced Wigner function $W_{\text{tr}_{-}(\rho)}$ is negative, the state must be entangled. This criterion is a special case of our lower bound when choosing $(x'=x_0-x,p'=p_0-p)$, where $x_0=\sqrt{2} x_{+}$ and $p_0=\sqrt{2} p_{+}$ are constants. This is because
\begin{equation}
\begin{aligned} \label{eq:WignerBS}
W_{\text{tr}_{-}(\rho)}\left(x_{+}, p_{+}\right) =& \iint_{-\infty}^{\infty} W\left(\frac{x_{+}+x_{-}}{\sqrt{2}}, \frac{p_{+}+p_{-}}{\sqrt{2}}, \frac{x_{+}-x_{-}}{\sqrt{2}}, \frac{p_{+}-p_{-}}{\sqrt{2}}\right) \dd x_{-} \dd p_{-} \\
=& 2 \iint_{-\infty}^{\infty} W\left(x_{-}, p_{-},-x_{-}+\sqrt{2} x_{+},-p_{-}+\sqrt{2} p_{+}\right) \dd x_{-} \dd p_{-} \\
=& 2 \iint_{-\infty}^{\infty} W\left(x,p,x',p'\right) \dd x \dd p,
\end{aligned}
\end{equation}
where the final line corresponds to our expression. This transformation has also been noted in~\cite{CertifiableZaw2024}.

Reference~\cite{CertifiableZaw2024} presents a more general entanglement criterion based on reduced states. The author demonstrates that negativity of the reduced Wigner function in mode $a_+$ implies partial transpose negativity. This approach extends previous work~\cite{DynamicsJayachandran2023} by incorporating additional local unitaries before and after passing the two modes through a balanced beam-splitter, corresponding to symplectic transformations of the coordinate system. Wigner negativity in one of the beam-splitter output modes indicates entanglement. This criterion is mathematically equivalent to our lower bound. Additionally, we provide upper bounds to the reduced Wigner function, further complementing the study of the relationship between this method based on the Wigner function and partial transposition.

\section*{VII. Comparison with Bell inequality and EPR steering criterion based on Wigner functions}

Compared to entanglement, EPR steering and Bell nonlocality are stronger forms of quantum correlations, and are thus more challenging to be prepared and detected in experiments. Some criteria based on Wigner functions have been proposed to detect EPR steering and Bell nonlocality in bipartite systems. In this section, we will compare the quantum correlations captured by different criteria for TMST, Werner and dephased cat states.

\subsection*{A. EPR criterion based on pseudospin measurements}

In \cite{KogiasHierarchy2015,YuPRA2017}, it is proposed the EPR steering criterion in terms of pseudospin measurements
\begin{align} \label{eq:YuEPRcriterion}
\mathcal{M} &= \langle \hat{\Pi}_A^x \otimes \hat{\Pi}_B^x \rangle ^2 + \langle \hat{\Pi}_A^y \otimes \hat{\Pi}_B^y \rangle ^2 + \langle \hat{\Pi}_A^z \otimes \hat{\Pi}_B^z \rangle ^2 \leq 1.
\end{align}
These pseudospin operators $(\hat{\Pi}^x,\hat{\Pi}^y,\hat{\Pi}^z)$ are closely associated to the Wigner representation \cite{RevzenPRA2005}
\begin{align}
W_{\Pi^x} (x,p) &=\text{sgn}(x), \nonumber  \\
W_{\Pi^y} (x,p) &=-\delta(x) \mathcal{P}\frac{1}{p}, \\
W_{\Pi^z} (x,p) &=-\pi \delta(x) \delta(p),
\end{align}
where $\mathcal{P}$ is the principal value. Therefore, expectation values of pseudospin operators can be directly obtained from the Wigner function as
\begin{align}
\langle \hat{\Pi}_A^j \otimes \hat{\Pi}_B^k \rangle =\frac{1}{(2\pi)^2} \int d^4 \mathbf{\xi} W_\rho(\mathbf{\xi}) W_{\Pi_A^j} (x_A,p_A) W_{\Pi_B^k} (x_B,p_B) ,
\end{align}
with $\mathbf{\xi}=(x_A,p_A,x_B,p_B)^T$.

\begin{figure}[t]
    \begin{center}
	\includegraphics[width=100mm]{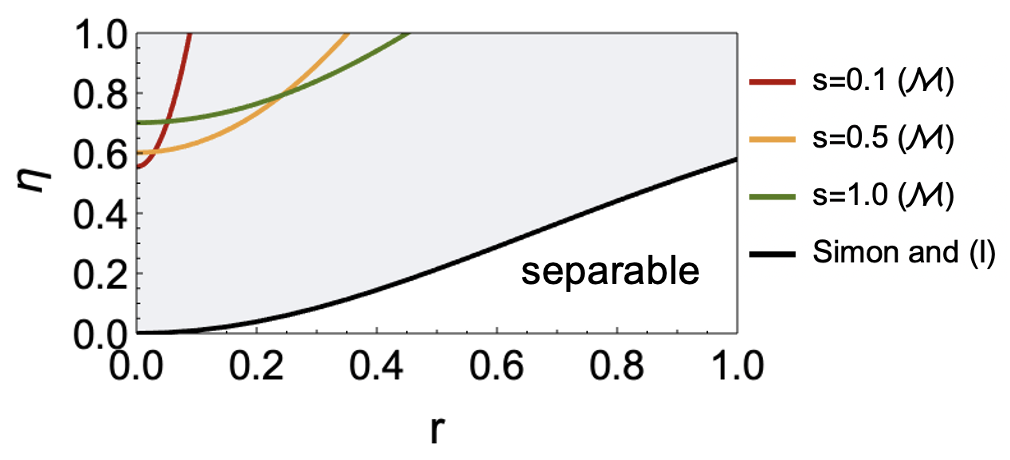}
	\end{center}
    \caption{Comparison between EPR steering criterion $\mathcal{M}$ and entanglement criterion \eqref{crit1} and Simon's for TMST states. Both criterion \eqref{crit1} and Simon's provide a tight bound for entanglement (black), which is the line $\eta=\tanh^2{(r)}$. $\mathcal{M}$ reveals EPR steering only for the states above red, yellow and green lines, which correspond to different squeezing parameter $s$.  }
    \label{Fig:SMEPRTMST}
\end{figure}

In Fig.~\ref{Fig:SMEPRTMST}, we compare EPR steering criterion $\mathcal{M}$ to criterion \eqref{crit1} and Simon's for TMST states. The black line represents the tight bound for entanglement, which can be provided by both criterion \eqref{crit1} and Simon's. $\mathcal{M}$ reveals EPR steering for TMST states with different $s$ (the region above red, yellow and green lines). Since entanglement is necessary but not sufficient for steering, criterion~\eqref{crit1} captures a wider region than the EPR steering criterion $\mathcal{M}$.

\subsection*{B. CHSH-type Bell inequality based on Wigner functions}

Reference \cite{BanaszekPRL1999} proposed the CHSH-type Bell inequality in terms of Wigner function measurements
\begin{align} \label{eq:BellWigner}
\mathcal{B}=| \langle \hat{P}_A (\alpha_A) \hat{P}_B (\alpha_B) \rangle + \langle \hat{P}_A (\alpha_A') \hat{P}_B (\alpha_B) \rangle + \langle \hat{P}_A (\alpha_A) \hat{P}_B (\alpha_B') \rangle - \langle \hat{P}_A (\alpha_A') \hat{P}_B (\alpha_B') \rangle  |\leq 2,
\end{align}
where $\hat{P}(\alpha) = \hat{D}^\dagger (\alpha) (-1)^{\hat{n}} \hat{D}(\alpha)$ is the displaced parity operator. The operator $\hat{P}$ can be expressed in terms of the Wigner function as $\langle \hat{P}_A(\alpha_A)\hat{P}(\alpha_B) \rangle= \frac{\pi^2}{4}W_{AB}(\alpha_A,\alpha_B)$.

\begin{figure}[H]
    \begin{center}
	\includegraphics[width=80mm]{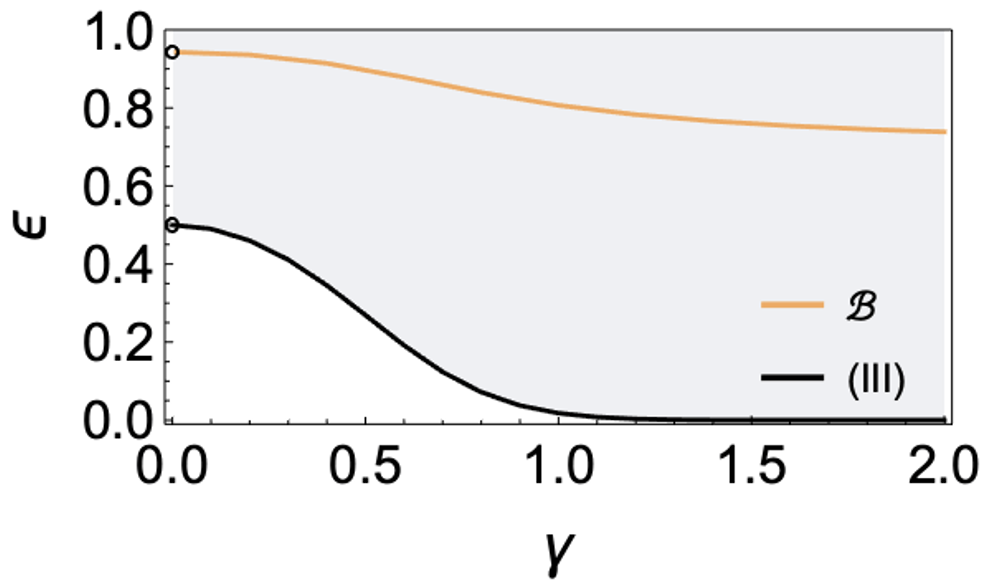}
	\end{center}
    \caption{ We compare the minimum $\epsilon$ for violating the Bell inequality \eqref{eq:BellWigner} and criterion \eqref{crit3} for the dephased cat state $\rho(\psi^-)$.}
    \label{Fig:SMBellCat}
\end{figure}

We have checked numerically that the near-optimal violation occurs when Alice's and Bob's displacement values are two identical imaginary numbers, \ie $Im(\alpha_A)=Im(\alpha_B), Im(\alpha'_A)=Im(\alpha'_B)$. For Werner states, the minimum parameters to violate Bell inequality in Eq.~\eqref{eq:BellWigner} are $\epsilon_{\min}\approx 0.9146$ for $\rho(\Phi^+)$ and $\epsilon_{\min}\approx 0.8919$ for $\rho(\Psi^+)$, respectively, which are much larger than the tight bound for entanglement ($\epsilon_{\min}=1/3$). For the dephased cat states $\rho(\psi^-)$, the comparison between Bell inequality \eqref{eq:BellWigner} and the entanglement criterion ~\eqref{crit3} is presented in Fig.~\ref{Fig:SMBellCat}. Since entanglement is necessary but insufficient for Bell nonlocality, criterion~\eqref{crit3} captures a wider range of entangled states (above the black line) than the Bell inequality~\eqref{eq:BellWigner} (above the yellow line).

\section*{VIII. Detailed analysis of some exemplary quantum states}
\subsection*{A. two-mode squeezed thermal states}
We start with two-mode squeezed thermal (TMST) states to investigate the applicability of our criteria to Gaussian states. An arbitrary TMST state can be generated by a specific scheme in \cite{YuPRA2017,MartinPRD2017}. A single-mode phase-insensitive Gaussian channel is applied to one of the modes of a two-mode squeezed vacuum state with squeezing parameter $s$. The channel is decomposed into a quantum-limited attenuator with transmissivity $\eta$ followed by a quantum limited amplifier with gain $\cosh^2 r$, introducing loss and noise to the states, respectively. 

The covariance matrix $V$ of the resulting TMST state has the form as Eq.~\eqref{eq:CMStandardForm}, with parameters
\begin{align}
n &= \eta \cosh^2{(r)} \cosh{(2s)} +(1-\eta) \cosh^2{(r)}+\sinh^2{(r)}, \nonumber \\
m &= \cosh{(2s)},  \\
c_1 &= -c_2=\sqrt{\eta} \cosh{(r)} \sinh{(2s)} \nonumber.
\end{align}
The Wigner functions of Gaussian states can be written as functions of the covariance matrix 
\begin{align}\label{eq:Wigner}
W = \frac{e^{-\frac{1}{2} \left( \vec{x}-\vec{\xi} \right)^T V^{-1} \left( \vec{x}-\vec{\xi} \right) }}{ (2\pi)^N \sqrt{\det V }},
\end{align}
where, $N$ is the number of modes, $\vec{x}=(x_1,p_1,\cdots,x_N,p_N)^T$ is a phase-space coordinate vector, and $\vec{\xi}$ is the displacement vector. We only consider the two-mode case ($N=2$) in our work. The analytical expressions of Wigner function for TMST states is obtained
\begin{align}
&W_{AB}(s,\eta,r) = \frac{1}{4\pi^2} \frac{1}{ \left( \cosh^2{(r)} \left( \eta-(\eta-1)\cosh{(2s)} \right)+\cosh{(2s)} \sinh^2{(r)} \right)} \times \notag\\
&e^{ -\frac{ -\left( \left( x_B^2+p_B^2\right) \left( \eta+(\eta-2)\cosh{(2r)} \right)  \right) + \left( 2\left(x_A^2+p_A^2\right)+\left(x_B^2+p_B^2 \right)\eta+ \left(x_B^2+p_B^2 \right) \eta \cosh{(2r)} \right) \cosh{(2s)} +4(p_A p_B-x_A x_B) \sqrt{\eta}\cosh{(r)}\sinh{(2s)}  }{ 2\left( 2\eta \cosh^2{(r)}-(\eta+(-2+\eta)\cosh{(2r)} )\cosh{(2s)} \right) } } \label{eq:TMSTWigner}
\end{align}

In the simplest case of two-mode squeezed vacuum (TMSV) states, \ie $\eta=1, r=0$, we find that under the optimal transformation $(x'=x,p'=-p)$ and the optimal parameter $\theta=\pi/4$, both criteria~(\ref{crit1},\ref{crit2}) can detect entanglement as long as $s>0$. For example, criterion~\eqref{crit1} can be expressed as
\begin{align}
\iint_{-\infty}^{\infty}  W_{AB}\left(\frac{\sqrt{2}}{2}x,\frac{\sqrt{2}}{2}p,\frac{\sqrt{2}}{2}x,-\frac{\sqrt{2}}{2}p \right) \dd x \dd p =\frac{e^{2s}}{2\pi}\leq \frac{1}{2\pi},
\end{align}
whose violation for $s>0$ indicates that the entanglement of TMSV states is detected.

Then let's analyze the most general case, \ie $1>\eta>0,r>0$, where Simon's criterion is still a necessary and sufficient conditions for detecting entanglement~\cite{SimonPRL2000}. We have proved in SM Sec.~V that our criterion~(\ref{crit1}) is  necessary and sufficient for Gaussian states, so criterion~(\ref{crit1}) provides a tight bound for detecting the entanglement for TMST states (See Fig.~\ref{Fig2_TMSTent} in the main text). 

Although useful for some states, criterion~(\ref{crit2}) no more provides a tight bound for all TMST states. Under the optimal linear transformation $(x'=x,p'=-p)$, the optimal parameter $\theta_{\text{opt}}$ in criterion~(\ref{crit2}) can be expressed as a function of $(s,\eta,r)$
\begin{align}
\theta_{\text{opt}}= \arctan \sqrt{\frac{-(8\pi\cosh{(2s)})}{ 4\pi\eta+4\pi(\eta-2)\cosh{(2r)} -8\pi\eta\cosh^2{(r)}\cosh{(2s)} }},
\end{align}
so that criterion~(\ref{crit2}) is reexpressed as
\begin{align}
\frac{ \sqrt{ -\frac{2\cosh{(2s)}}{ \eta+(\eta-2)\cosh{(2r)} -2\eta \cosh^2{(r)} \cosh{(2s)} } } }{ \cosh^2{(s)}+\sinh^2{(s)}-\sqrt{2\eta} \cosh{(r)} \sinh{(2s)} \sqrt{ -\frac{\cosh{(2s)}}{ \eta+(\eta-2) \cosh{(2r)} -2\eta \cosh^2{(r)} \cosh{(2s)}}  }. }-1\leq 0.
\end{align}
We illustrated the performance of criterion~(\ref{crit2}) in Fig.~\ref{Fig2_TMSTent} of the main text, for the values of $s=0.1, 0.5, 1.0$.

\subsection*{B. Werner states}

\begin{figure}[t]
    \begin{center}
	\includegraphics[width=120mm]{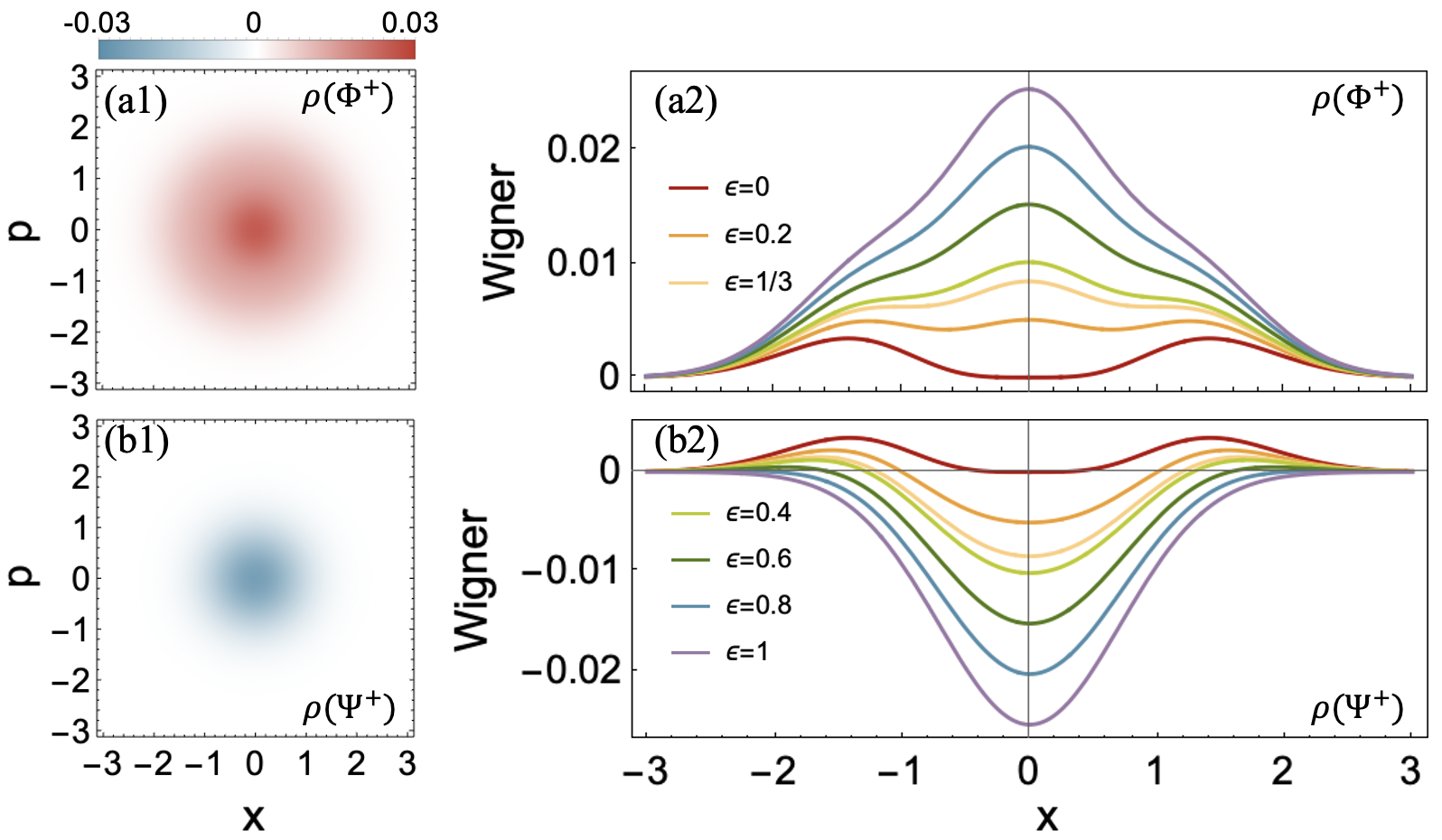}
	\end{center}
    \caption{Wigner distributions of Werner states. For the Werner state $\rho(\Phi^+)$, (a1) the Wigner functions $W_{AB}(x,p,x,-p)$ at $\epsilon=1$, (a2) the horizontal line-cuts when $p=0$, \ie $W_{AB}(x,0,x,0)$, for different $\epsilon$. It is shown that $W_{AB}(x,p,x,-p)$ is positive over the entire space for any $\epsilon$. For the Werner state $\rho(\Psi^+)$, (b1) the Wigner functions $W_{AB}(x,p,-x,-p)$ at $\epsilon=1$, (b2) the horizontal line-cuts of $W_{AB}(x,0,-x,0)$, for different $\epsilon$.}
    \label{Fig:SMWerner}
\end{figure}

Werner states are of the form
\begin{align}
\rho &= \epsilon |\psi \rangle \langle \psi | + \frac{1-\epsilon}{4} \mathbbm{1} ,
\end{align}
where $|\psi\rangle$ is a pure state and $\epsilon \in [0,1]$ is a parameter determining the purity of $\rho$. In this section, we will consider two Bell states: (1) $|\psi\rangle=|\Phi^+\rangle \equiv  \left( |00\rangle +|11\rangle \right) / \sqrt{2}$ and (2) $|\psi\rangle =|\Psi^+\rangle \equiv  \left( |01\rangle +|10\rangle \right) / \sqrt{2}$. 

First, let's consider the Werner state $\rho (\Phi^+)$, whose Wigner function is
\begin{align} \label{eq:WignerWerner0011}
W_{AB}(x_A,p_A,x_B,p_B) 
&= \frac{1}{16\pi^2} e^{-\frac{1}{2}\left(x_A^2+p_A^2+x_B^2+p_B^2 \right)} \Big[ (1+\epsilon)\left(x_A^2 x_B^2+x_A^2 p_B^2+ p_A^2 x_B^2+p_A^2 p_B^2 \right) \nonumber \\
& \quad +4\epsilon (x_A x_B-p_A p_B) -2\epsilon\left(x_A^2+p_A^2+x_B^2+p_B^2 \right) +4\epsilon \Big].
\end{align}
In Fig.~\ref{Fig:SMWerner}(a), Wigner distribution for $\rho (\Phi^+)$ under the linear transformation $(x'=x,p'=-p)$ is illustrated, which is non-negative for any $\epsilon$. With $\theta=\pi/4$, the criteria~(\ref{crit1},\ref{crit2}) with the integration over the entire space are given by
\begin{align}
&\iint_{-\infty}^{\infty} W_{AB} \left( \frac{\sqrt{2}}{2}x, \frac{\sqrt{2}}{2}p, \frac{\sqrt{2}}{2}x, -\frac{\sqrt{2}}{2}p  \right) \mathrm{d} x \mathrm{d} p \nonumber \\
&=\iint_{-\infty}^{\infty} \abs{W_{AB} \left( \frac{\sqrt{2}}{2}x, \frac{\sqrt{2}}{2}p, \frac{\sqrt{2}}{2}x, -\frac{\sqrt{2}}{2}p  \right) } \mathrm{d} x \mathrm{d} p \nonumber \\
& =\frac{1+3\epsilon}{4\pi} \nonumber \\
& \leq \frac{1}{2\pi}.
\end{align}
Both criteria~(\ref{crit1},\ref{crit2}) are violated for $\epsilon>1/3$, revealing entanglement in $\rho(\Phi^+)$. This bound is equivalent with the known tight bound obtained by PPT criterion~\cite{SeparabilityHorodeckiPLA1996}.

For Werner state $\rho (\Psi^+)$, whose Wigner function is
\begin{align}\label{eq:WignerWerner0110}
W_{AB}(x_A,p_A,x_B,p_B) 
&= \frac{1}{16\pi^2} e^{-\frac{1}{2}\left(x_A^2+p_A^2+x_B^2+p_B^2 \right)} \Big[ (1-\epsilon)\left(x_A^2 x_B^2+x_A^2 p_B^2+ p_A^2 x_B^2+p_A^2 p_B^2 \right) \nonumber \\
& \quad +4\epsilon (x_A x_B+p_A p_B) +2\epsilon\left(x_A^2+p_A^2+x_B^2+p_B^2 \right) -4\epsilon \Big].
\end{align}
We find that the Wigner function $W_{AB} (x,p,-x,-p)$ is negative over the entire space at $\epsilon=1$. With $\theta=\pi/4$, criterion~\eqref{crit3} is expressed as
\begin{align}
\iint_{-\infty}^{\infty} W_{AB} \left(x,p,-x,-p\right) \dd x \dd p =\frac{1-3\epsilon}{8\pi} \geq 0.
\end{align}
Criterion~\eqref{crit3} reveals entanglement for $\rho(\Psi^+)$ when $\epsilon>1/3$, which coincides with the known tight bound obtained by PPT criterion~\cite{SeparabilityHorodeckiPLA1996}.

In conclusion, our criteria, which perform as well as PPT criterion, are necessary and sufficient conditions for the considered Werner states. In addition, we have analyzed some known entanglement criteria for better comparison in Fig.~\ref{Fig:SMtable} of the main text. It's found that Duan-Simon criteria~\cite{SimonPRL2000,DuanPRL} are unable to detect entanglement for any $\epsilon$, and the entanglement based on quantum Fisher information \cite{ManuelPRA2016} and Hillery-Zubairy criteria~\cite{HilleryPRL2006} are useful but not tight (the former reveals entanglement for both $\rho(\Phi^+)$ and $\rho(\Psi^+)$ when $\epsilon \gtrsim 0.65$, and the latter can only detect entanglement for $\rho(\Psi^+)$ when $\epsilon >\frac{\sqrt{5}-1}{2}$ ).

\subsection*{C. partially dephased two-mode cat states}
\begin{figure}[t]
    \begin{center}
	\includegraphics[width=\textwidth]{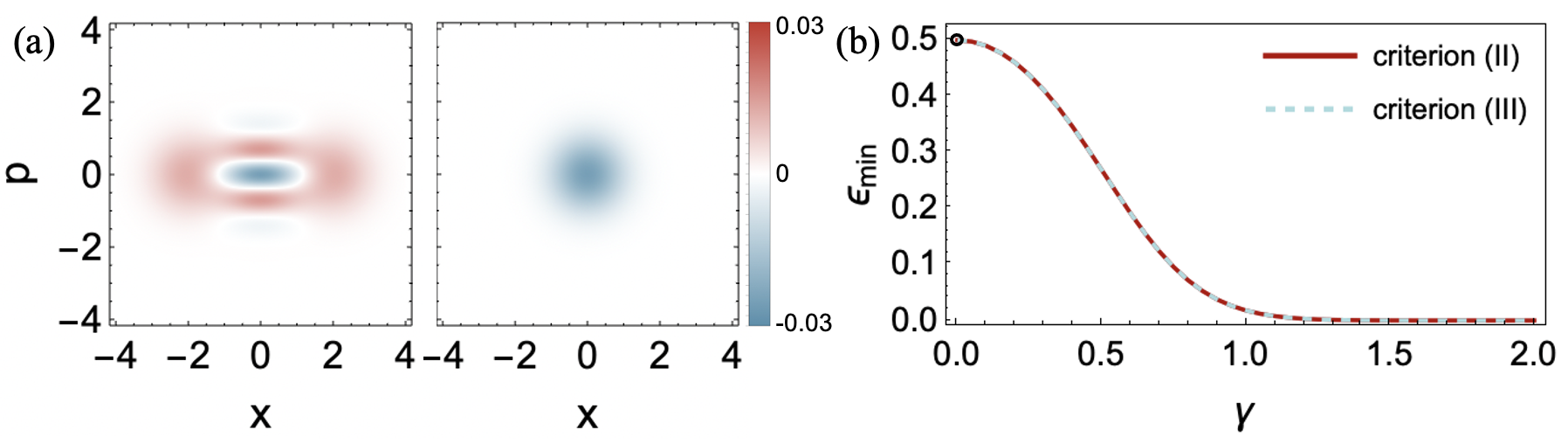}
	\end{center}
    \caption{Wigner distributions and entanglement detection of partially dephased two-mode cat state $\rho(\psi^-)$. (a) The Wigner function $W_{AB}(x,p,x,p)$ (left) and $W_{AB}(x,p,-x,-p)$ (right) at $\gamma=1,\epsilon=1$. (b) The minimum parameter $\epsilon_{\min}$ required to violate the criteria~(\ref{crit2},\ref{crit3}) as a function of $\gamma$. Note $\gamma=0$ gives a singular point. }
    \label{Fig:SMCatMinus}
\end{figure}

A dephased two-mode cat state is of the form
\begin{align}
\rho &=\epsilon |\psi\rangle \langle \psi|+\frac{(1-\epsilon)}{2} \left( |\gamma,\gamma\rangle \langle \gamma,\gamma|+|-\gamma,-\gamma\rangle \langle -\gamma,-\gamma| \right),
\end{align}
where $\epsilon\in[0,1]$ is a parameter specifying the dephasing, and $|\psi\rangle=|\psi^+\rangle \equiv \mathcal{N} \left( |\gamma, \gamma\rangle+|-\gamma, -\gamma\rangle \right)$ is a two-mode cat state with coherent states $|\gamma\rangle$ and normalization constant $\mathcal{N}$. 

The Wigner distribution for $\rho(\psi^+)$ is written as
\begin{align}
W_{AB}(x_A,p_A,x_B,p_B) &= \frac{ e^{-\frac{1}{2} \left(x_A^2+p_A^2+x_B^2+p_B^2+8\gamma^2 \right) } }{4\pi^2\left( e^{4\gamma^2}+1 \right) } \nonumber \\
& \times \left( e^{8\gamma^2}\epsilon \cos{(2\gamma(p_A+p_B))} +\left(e^{4\gamma^2}-\epsilon+1 \right) \cosh{(2\gamma(x_A+x_B))}  \right).
\end{align}
Under the transformation $(x'=x,p'=-p)$ and $\theta=\pi/4$, the resulting Wigner function is positive over the entire space for any $\epsilon$. Criterion~(\ref{crit1}) can be calculated as
\begin{align}
\iint_{-\infty}^{\infty} W_{AB} \left( \frac{\sqrt{2}}{2}x,\frac{\sqrt{2}}{2}p,\frac{\sqrt{2}}{2}x,-\frac{\sqrt{2}}{2}p \right) = \frac{1+\epsilon \tanh{(2\gamma^2)}}{2\pi} \leq \frac{1}{2\pi},
\end{align}
which is violated for any $\gamma>0$ and $\epsilon\in(0,1]$. Criterion~(\ref{crit2}) will yield the same results if an integration over the entire space is considered. We conclude that both criteria~(\ref{crit1},\ref{crit2}) can provide tight bounds for $\rho(\psi^+)$ (see also Fig.~\ref{Fig3_Catent} in the main text). 

We also consider the dephased cat states $\rho(\psi^-)$ with $|\psi^-\rangle \equiv \mathcal{N}' \left( |\gamma, \gamma\rangle - |-\gamma, -\gamma\rangle \right)$, whose Wigner functions are
\begin{align}\label{eq:Wignerdephasedcat}
W_{AB}(x_A,p_A,x_B,p_B) &= \frac{ e^{-\frac{1}{2} \left(x_A^2+p_A^2+x_B^2+p_B^2+8\gamma^2 \right) } }{4\pi^2\left( e^{4\gamma^2}-1 \right) } \nonumber \\
& \times \left( -e^{8\gamma^2}\epsilon \cos{(2\gamma(p_A+p_B))} +\left(e^{4\gamma^2}+\epsilon-1 \right) \cosh{(2\gamma(x_A+x_B))}  \right).
\end{align}
Since the Wigner function under the linear transformation $(x'=-x, p'=-p)$ is negative at $\epsilon=1$ (see Fig.~\ref{Fig:SMCatMinus}(a)), we calculate criterion~\eqref{crit3} 
\begin{align} \label{eq:TMCSCriterion3}
\iint_{-\infty}^{\infty} W_{AB} \left(x,p,-x,-p\right) \dd x \dd p = \frac{ e^{-4\gamma^2} \left( 1-\left(1+ e^{4\gamma^2} \right)\epsilon \right) }{4\pi}.
\end{align}
For separable states this quantity must be non-negative, so entanglement is revealed when $\epsilon> 1 / (1+e^{4 \gamma^2 })$. We also calculate criterion~\eqref{crit2} numerically, under the transformation $(x'=x,p'=p)$ and $\theta=\pi/4$. As illustrated in Fig.~\ref{Fig:SMCatMinus}(b) , the minimum $\epsilon_{\min}$ for violating criterion~\eqref{crit2} coincides with the one for violating criterion~\eqref{crit3}. 

In Fig.~\ref{Fig:SMtable} of the main text, as a summary, we compare the entanglement detection capabilities of our criteria~(\ref{crit1},\ref{crit2},\ref{crit3}) to Duan-Simon~\cite{SimonPRL2000,DuanPRL}, Hillery-Zubairy~\cite{HilleryPRL2006}, QFI~\cite{ManuelPRA2016},  and PPT~\cite{PeresPRL1996,SeparabilityHorodeckiPLA1996} criteria, respectively. For each of the considered states, at least one criterion out of the family (\ref{crit1},\ref{crit2},\ref{crit3}) can provide a tight bound. Note that, different from QFI criterion~\cite{ManuelPRA2016} and PPT criterion~\cite{PeresPRL1996,SeparabilityHorodeckiPLA1996} which require full information of the density matrix, the criteria (\ref{crit1},\ref{crit2},\ref{crit3}) only needs partial information determined by the transformation $(x',p')$.

\section*{IX. Error analysis}
\subsection*{A. Discretization}

The entanglement criteria we proposed are written as integrals over the (continuous) joint Wigner function. 
In realistic experiments, however, the Wigner function is sampled over a finite set of points in phase space through displaced parity measurements.
For this reason, integrals need to be approximated by summations over the discrete dataset available. 
Consequently, the sampling resolution, i.e. the effective phase-space pixel size, introduces a finite-sampling error that we quantify below.

We first benchmark criterion~\eqref{crit2} using the pure cat states  $|\psi\rangle =\mathcal{N} \left( |\gamma,\gamma\rangle+|-\gamma,-\gamma\rangle \right)$ with the size $\gamma=2$. 
In the ideal case, the amount of violation is expressed by the integral over a finite region $R$, 
\begin{align}
I &= \iint_{R} \abs{ W_{AB}(x,p,x,-p) } \dd x \dd p -\frac{1}{4\pi}.
\end{align}

\begin{figure}[t]
    \begin{center}
	\includegraphics[width=\textwidth]{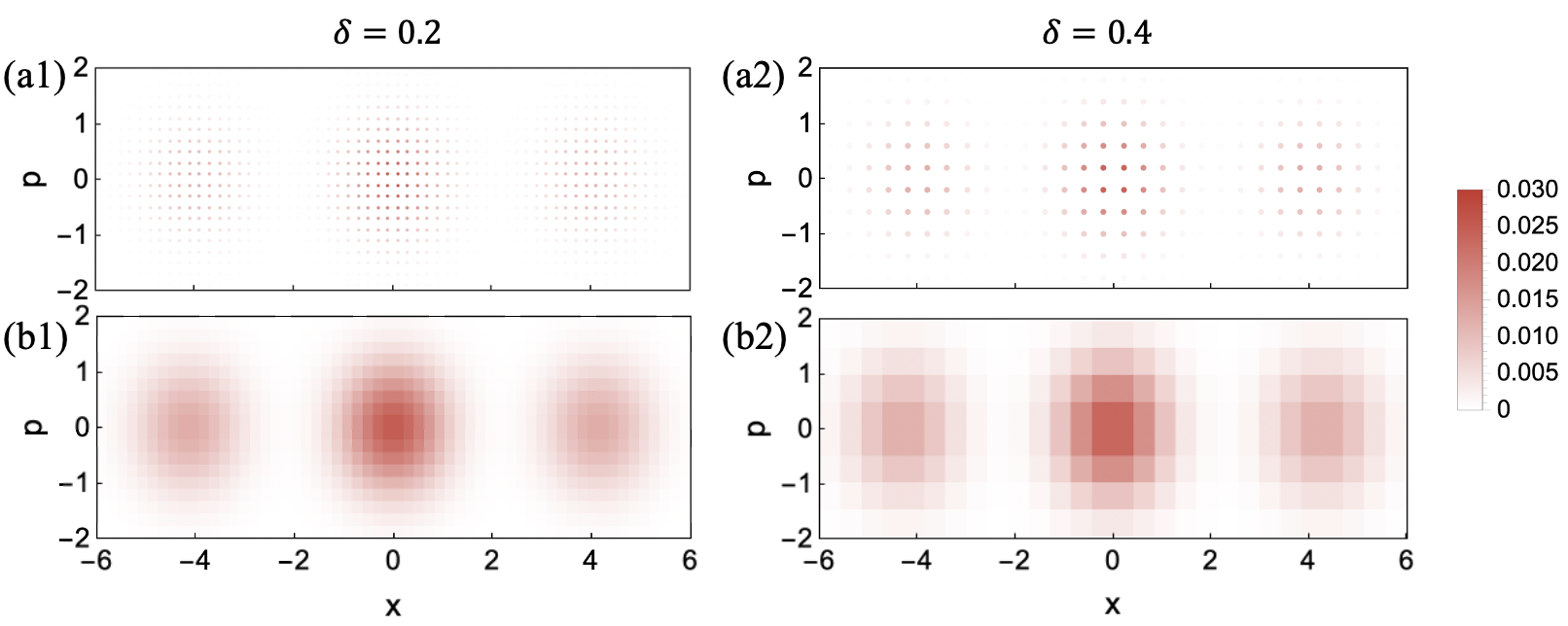}
	\end{center}
    \caption{ The effect of sampling resolution $\delta$ on the approximation of joint Wigner functions $W_{AB}$. We consider two-mode cat states $|\psi\rangle$ with $\gamma=2$, and compare $\delta=0.2$ and $\delta=0.4$.  (a1)(a2) The discrete points $W_{AB}(x_j,p_k,x_j,-p_k)$. (b1)(b2) These discrete points are processed by assigning their values to the corresponding region $\delta^2$.  }
    \label{Fig:SMFiniteSampling}
\end{figure}

In a practical scenario, we imagine that the experiment can sample $W_{AB}$ on a square grid of spacing $\delta$ over the finite region $R$. 
Specifically, we define the discrete points along the $x$-axis as $x_{j}=x_{\text{min}}+j\delta$ with $j=0,1,2,\cdots,(x_{\text{max}} -x_{\text{min}})/\delta$, and analogously, the discrete points $p_k$ along the $p$-axis, yielding the discrete Wigner functions $W_{AB}(x_j,p_k,x_j,-p_k)$. 
In Fig.~\ref{Fig:SMFiniteSampling}, we compare the discrete Wigner functions $W_{AB}(x_j,p_k,x_j,-p_k)$ under sampling resolution $\delta=0.2$ and $\delta=0.4$, showing that the Wigner functions become coarser as $\delta$ grows.

\begin{figure}[t]
    \begin{center}
	\includegraphics[width=90mm]{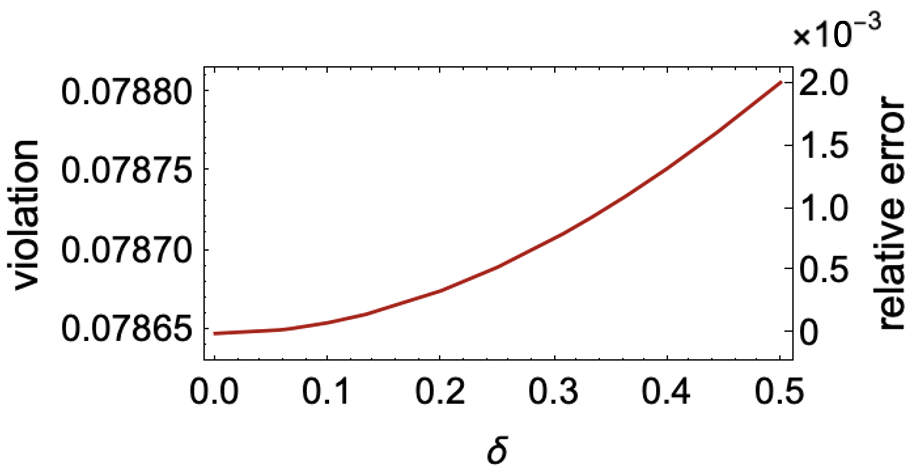}
	\end{center}
    \caption{ The violation and relative error as a function of sampling resolution $\delta$.  }
    \label{Fig:SMReletiveError}
\end{figure}

In this finite-sampling situation, the violation values are computed from the discrete Wigner function as
\begin{align}
I_{D} &= \sum_{j,k} \abs{ W_{AB} \left( x_j, p_k, x_j, -p_k \right) } \delta^2 -\frac{1}{4\pi}.
\end{align}
Fig.~\ref{Fig:SMReletiveError} illustrates the amount of violation $I_D$ and relative error $\epsilon_{\text{rel}}=|I_D-I|/|I|$ as a function of the sampling resolution $\delta$. It is found that our criterion is robust against finite sampling under realistic discrete approximations, with the relative error as small as $\epsilon_{\text{rel}} \approx 2\times 10^{-3}$ even for a coarse sampling $\delta=0.5$. 

We observe from Fig.~\ref{Fig:SMFiniteSampling} that the Wigner function $W_{AB}(x_j,p_k,x_j,-p_k)$ for pure cat states $|\psi\rangle$ remains non-negative over the entire phase space. 
Similar non-negative distribution can also be found in $W_{AB}(x_j,p_k,x_j,-p_k)$ for two-mode squeezed vacuum (TMSV) states and Bell states $|\Phi^+\rangle$. 
In both these cases, we restrict the sampling region $R$ to $x,p\in[-3,3]$, and observe the violation values increase slightly as $\delta$ grows. 
At $\delta=0.5$, the relative error is found to be $\epsilon_{\text{rel}} \approx 3 \times 10^{-4}$ for the TMSV states and $\epsilon_{\text{rel}} \approx 6 \times 10^{-4}$ for Bell states $|\Phi^+\rangle$.

\begin{figure}[t]
    \begin{center}
	\includegraphics[width=\textwidth]{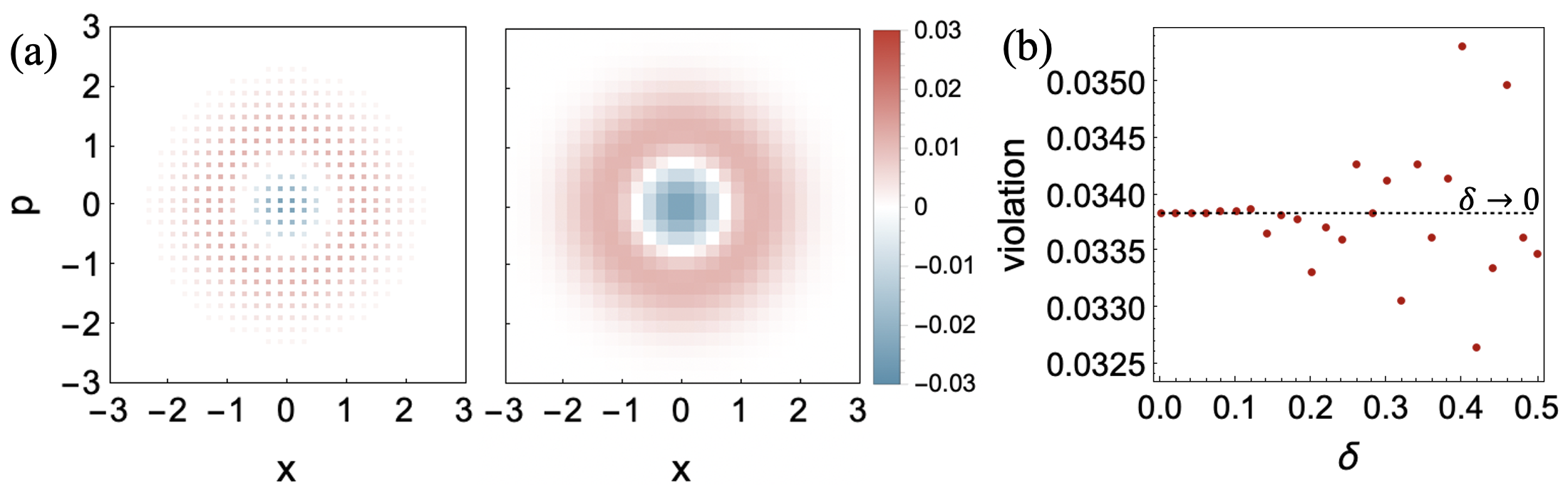}
	\end{center}
    \caption{ The effect of sampling resolution $\delta$ on pure Bell states $|\Psi^+\rangle$. (a) The discrete points $W_{AB}(x_j,p_k,x_j,p_k)$ at the sampling resolution $\delta=0.2$. (b) The violation $I_D$ changes with $\delta$, where the black dashed line represents $I$ in $\delta \rightarrow 0$ ideal case.  }
    \label{Fig:SMFSBell}
\end{figure}

On the other hand, for the Bell state $|\Psi^+\rangle$ the joint Wigner function $W_{AB}(x,p,x,p)$ in criterion~\eqref{crit2} involves both positive and negative values. Fig.~\ref{Fig:SMFSBell}(a) shows the discrete Wigner functions $W_{AB}(x_j,p_k,x_j,p_k)$ for a sampling resolution of $\delta=0.2$. Fig.~\ref{Fig:SMFSBell}(b) presents how the amount of violation changes with sampling resolution $\delta$. 
Contrary to the non-negative case, the change of violation is no longer monotonic with the sampling resolution $\delta$, reflecting the contribution of both positive and negative pixels. 
Within sampling resolution $\delta\in[0,0.5]$, we find the largest relative error $\epsilon_{\text{rel}} \approx 0.04$ at $\delta=0.4$.
For $\delta < 0.12$, the observed violation clearly approaches the ideal value given by $\delta\rightarrow 0$ to very good approximation. 

Criteria~(\ref{crit1},\ref{crit3}) involve integrals over the entire phase space. Since Wigner functions tend to zero at infinity, we can choose a sufficiently large region $R'$, where the contribution outside of $R'$ is negligible. 
The criteria can then be computed from a discrete set of measurements within this region.

\subsection*{B. How to measure the criteria}

Each individual displaced-parity measurement yields as outcome ±1, and averaging over many measurement repetitions provides the value of the Wigner function at the corresponding point in phase-space. For a total finite number of measurements $N$, let each measurement result be a random variable $X_i \in \{+1, -1\}$ with mean value $E[X_i] = \mu$ (which is proportional to $W(\alpha; \beta)$). The sample mean is
\begin{equation}
\hat{\mu}\left(X_i\right)=\frac{1}{N} \sum_{i=1}^N X_i .
\end{equation}
with variance 
\begin{equation}\label{suppeq:var}
\sigma^2(X_i) = \frac{1}{N^2} \sum_{i=1}^N \sigma^2(X_i) = \frac{1-\mu^2}{N} .
\end{equation}

To illustrate our approach, we consider the two-mode squeezed vacuum (TMSV) state with Wigner function $W(\alpha ; \beta)=\frac{4}{\pi^2} \exp\left[ -2 \cosh 2 r\left(|\alpha|^2+|\beta|^2\right)+2 \sinh 2 r\left(\alpha \beta+\alpha^* \beta^*\right) \right]$ where $r$ is the squeezing parameter. 
We evaluate the CHSH combination $\mathcal{B}$ as
\begin{equation}
\begin{aligned}
\mathcal{B} & =\Pi(0 ; 0)+\Pi(\sqrt{\mathcal{J}} ; 0)+\Pi(0 ;-\sqrt{\mathcal{J}})-\Pi(\sqrt{\mathcal{J}} ;-\sqrt{\mathcal{J}}) \\
& =1+2 \exp (-2 \mathcal{J} \cosh 2 r)-\exp \left(-4 \mathcal{J} e^{2 r}\right).
\end{aligned}
\end{equation}
According to Bell's theorem, classical correlations satisfy $-2 \leq \mathcal{B} \leq 2$.
For the TMSV state with fixed squeezing parameter $r$, the optimal displacement amplitude $\mathcal{J}_{\mathrm{opt}}(r)=\frac{2 r-\ln (\cosh 2 r)}{3 e^{2 r}-e^{-2 r}}$ maximizes the violation of the CHSH inequality \cite{BanaszekPRL1999,BanaszekPRA1998}. This choice yields
\begin{equation}
\mathcal{B}_{max}= \left( 2 \tanh (2 r)+1 \right) \exp \left(-\frac{4 e^{4 r} (2 r+\log (\text{sech}(2 r)))}{3 e^{4 r}-1}\right)+1.
\end{equation}
In the limit $r \rightarrow \infty$, this reduces to $\mathcal{J} e^{2 r}=\frac{1}{3} \ln 2$ and $\mathcal{B}_{max}=1+3 \times 2^{-4 / 3} \approx 2.19055$.

Note that $\Pi(0 ; 0)=1$, $\Pi(\sqrt{\mathcal{J}} ; 0)=\Pi(0 ;-\sqrt{\mathcal{J}})=\exp (-2 \mathcal{J} \cosh 2 r)$ and $\Pi(\sqrt{\mathcal{J}} ;-\sqrt{\mathcal{J}})=\exp \left(-4 \mathcal{J} e^{2 r}\right)$. When each of these four parity points is sampled with $N/4$ trials, the estimator variance becomes
\begin{equation}
\sigma^2(\widehat{\mathcal{B}})=\frac{8\left[1-e^{-4 \mathcal{J} \cosh 2 r}\right]+4\left[1-e^{-8 \mathcal{J} e^{2 r}}\right]}{N}.
\end{equation}
For a squeezing parameter $r=1$, the ideal $\mathcal{B}_{\text{max}}$ equals $2.1839$. To establish a figure of merit, we calculate the minimum number of measurements $N_{\text{min}}$ required for violations at different confidence levels. Substituting $\mathcal{J}_{\text{opt}}(r)$ into the above expression, we obtain for a $10$-standard-deviation (10$\sigma$) violation, $N_{\text{min}}=18632$; for 5$\sigma$, $N_{\text{min}}=4658$; and for 3$\sigma$, $N_{\text{min}}=1677$.

We now investigate the amount of data needed when using criteria (\ref{crit1},\ref{crit2}) to detect entanglement on the same state. Due to the state's symmetry, we set $\theta = \pi/4$, which reduces both criteria to
\begin{equation}
\iint_{-\infty}^{\infty} W_{A B}(x, p, x,-p) \mathrm{d} x \mathrm{~d} p \leq \frac{1}{4 \pi}
\end{equation}
In practice, measurements can only cover a finite phase-space region, which we take as a disk of radius $\alpha_0$ centered at the origin. Our criteria become
\begin{equation}
\langle \Pi_{\text{eff}} \rangle:=\frac{1}{\pi \alpha_0^2}\int_{x^2+p^2\leq \alpha_0^2} \langle \Pi_{A B}(x, p, x,-p) \rangle \mathrm{d} x \mathrm{~d} p \leq \frac{1}{\alpha_0^2}
\end{equation}
where $\Pi$ denotes the displaced parity measurement and $\langle \Pi_{A B}(x, p, x,-p)\rangle=e^{-e^{-2 r}  \left(p^2+x^2\right)}$. The quantity $\Pi_{\text{eff}}$ can be understood as the average outcome of displaced parity measurements within the region. In the following we show why it can be regarded effectively as a single-point measurement. The average outcome of displaced parity measurements is
\begin{equation}\label{suppeq:muG}
\mu=\frac{e^{2 r} \left(1-e^{\alpha_0^2 \left(-e^{-2 r}\right)}\right)}{\alpha_0^2}
\end{equation}

We propose evaluating the integral using a randomized (or “scanning”) measurement protocol over the region of interest.  In conventional approaches, measuring the Wigner function at a fixed point $(x, p, x, -p)$ requires applying the displacement operator $D(x, p) \otimes D(x, -p)$ followed by multiple measurements of the joint parity operator $(-1)^{\hat{n}_a + \hat{n}_b}$. This produces a sequence of $\pm1$ outcomes whose mean is proportional to the Wigner function at that point. Since our criterion requires only the mean value over the 2D disk of radius $\alpha_0$, rather than the complete profile of the 2D Wigner function slice, we can vary the displacement continuously during each parity measurement to uniformly sample the integration region. 

We only need to make sure that the displacements are sampled uniformly over the desired region. These measurements are statistically equivalent to evaluating the Wigner function at a single point. We now show that this randomized measurement scheme significantly reduces the required number of measurements. From Eqs.~(\ref{suppeq:var},\ref{suppeq:muG}), the variance is
\begin{equation}
\sigma^2=\frac{1-\frac{e^{4 r} \left(1-e^{\alpha_0^2 \left(-e^{-2 r}\right)}\right)^2}{\alpha_0^4}}{N}.
\end{equation}
To achieve a violation of our inequality by at least $10$ standard deviations, we solve $\Pi_{\text {eff }}-\frac{1}{\alpha_0^2} \geq 10 \sigma$ and find the minimum total number of measurements $N_{\text{min}}=136$ with optimal region radius $\alpha_0=1.70583$. 
Notably, this is two orders of magnitude fewer points than the one required to attain a comparable violation of the CHSH inequality. 
For 5$\sigma$: $N_{\text{min}}=34,\alpha_0=1.70583$; for 3$\sigma$: $N_{\text{min}}=13,\alpha_0=1.71584$.

Another notable distinction from CHSH inequality tests is that while the CHSH violation has an upper bound of $2\sqrt{2}$, the violation of our entanglement criteria has no such upper limit. For states with squeezing approaching infinity, the violation of our criteria can grow unboundedly. So the entanglement detection becomes more robust to errors: if the experiment is limited by some systematic or statistical errors in Wigner integration, one can compensate by generating larger squeezing to yield a larger violation value over the separable threshold. 

The randomized measurement protocol can be implemented across different experimental platforms. In microwave cavities, mechanical oscillators, and ion-trap systems, computer-controlled experiments allow straightforward implementation of displacement scanning functions. The displacements can be distributed either randomly or deterministically over the measurement region. For uniform sampling on a disk, deterministic distributions such as the Fermat spiral provide an efficient approach (see, e.g., \cite{VogelSunflower1979}). The key requirement is ensuring uniform coverage of the selected phase-space region through appropriate programming of the displacement operations.

\subsection*{C. Additional comments on error analysis and number of measurements}

\begin{figure}[t]
    \begin{center}
	\includegraphics[width=0.9\columnwidth]{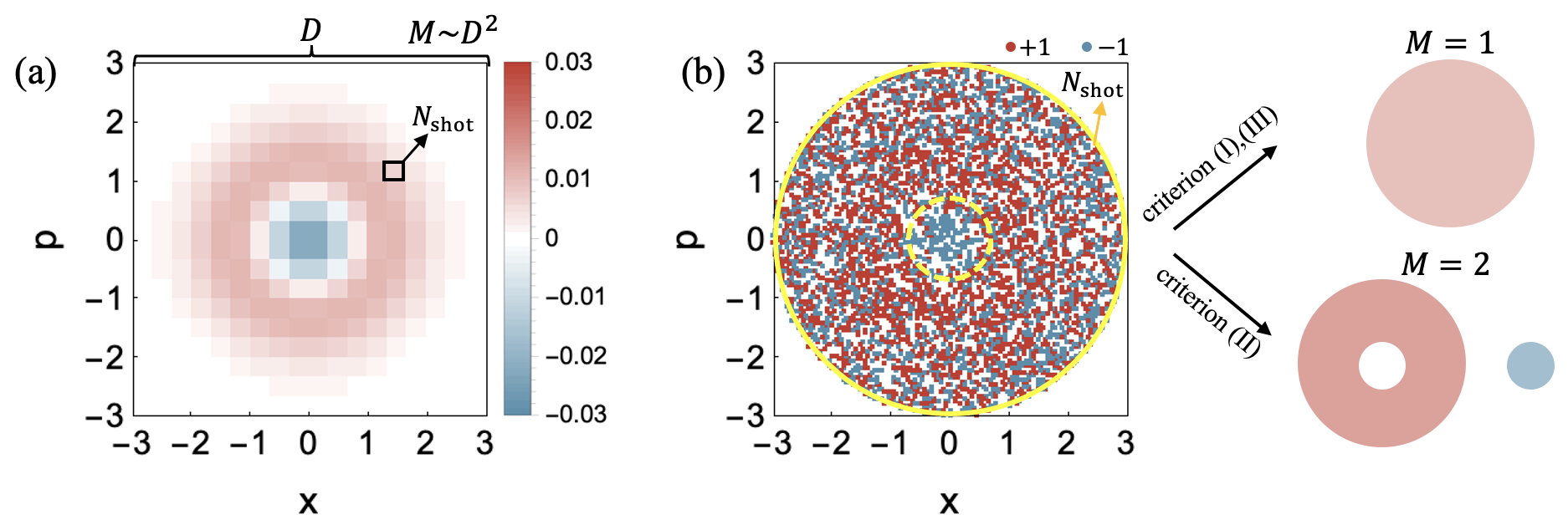}
	\end{center}
    \caption{Illustration of (a) standard detection and (b) randomized detection. In (a) standard detection, the phase space is discretized into $M\sim D^2$ samples, and $N_{\text{shot}}$ repeated shots are required for each sample. In (b) randomized detection, $N_{\text{shot}}$ random shots are performed within the chosen space (yellow solid circle). Criteria (I) and (III) require the average values over all statistics, which is equivalent to $M=1$ expectation value. Criterion (II) requires the average values over positive and negative regions, respectively, corresponding to $M=2$ expectation values. In both methods, the amount of total data is $M\cdot N_{\text{shot}}$.}
    \label{Fig_IllustrationTwoDetections}
\end{figure}

In the preceding two subsections, we analyzed errors that are independent of specific experimental apparatus details but may still play a dominant role in practical implementations. We now consider the overall uncertainty including apparatus-level contributions. We characterize the combined uncertainty from random and systematic errors as $\sigma_{\mathrm{tot}}=\sqrt{\sigma_{\mathrm{stat}}^2+\sigma_{\mathrm{syst}}^2}$. According to the discussion in the previous subsections, if grid discretization is employed to evaluate the two-dimensional integral, the associated error can be classified as systematic, since the bias originates from the finite grid spacing, i.e., the discretization approximation due to non-infinitesimal $\delta$. In contrast, if randomized sampling measurements are used to estimate the integral over the integration region, the error can be classified as statistical, as it arises from finite repetitions and decreases with increasing sample size. Beyond these considerations, the experimental apparatus itself may introduce additional systematic errors, such as parity readout errors and displacement calibration errors.

We now elaborate on how to compare measurement counts. In typical direct Wigner function measurement experiments, the total number of measurements can be written as $N_{\mathrm{tot}}=M \cdot N_{\mathrm{shot}}$, where $M$ is the number of sampled displacement points and $N_{\mathrm{shot}}$ is the number of repeated shots at each displacement point. For full tomography of a two-mode quantum state, if the effective truncation dimension of each mode is $D$, the required number of measurement settings $M$ typically scales as $D^4$. In contrast, under our grid discretization scheme from the previous subsections, we effectively reconstruct a two-dimensional phase-space distribution, so $M$ scales as $D^2$. The advantage of this approach is that it provides the most straightforward way to measure and compute the various criteria when no prior knowledge of the quantum state's principal distribution region in phase space is available. However, when the experimentalist knows in advance which quantum state will be prepared and has prior information about the principal phase-space distribution region, employing randomized measurements can substantially reduce the required $M$. 

Specifically, in the randomized measurement protocol, the number of measurement rounds can be taken as $M=1$ for criteria~\eqref{crit1} and~\eqref{crit3}, and $M=2$ for criterion~\eqref{crit2}. This is because we employ randomized sampling: within a single measurement round, $N_{\mathrm{shot}}$ randomly displaced sampling points are included, with each sampling point measured only once (yielding a single $\pm 1$ outcome) rather than repeating $N_{\mathrm{shot}}$ times at the same point. Thus, one measurement round still corresponds to $N_{\mathrm{shot}}$ shots. Criteria~\eqref{crit1} and~\eqref{crit3} involve linear integrals, and the required data is essentially a single mean value over the two-dimensional phase space. Estimating this mean via randomized sampling is statistically equivalent to directly measuring the Wigner function at a single point. Criterion~\eqref{crit2}, on the other hand, requires partitioning the distribution into positive and negative regions based on prior knowledge, and performing mean-value measurements separately over each region, thus necessitating two measurement rounds. Even if the prior partitioning is imprecise or the sign of certain regions is misjudged, this will only lower the computed value of criterion~\eqref{crit2} rather than increase it, thereby weakening the detection capability without falsely indicating entanglement.

\end{document}